\documentclass[12pt]{article}
\usepackage{amssymb,amsthm}
\usepackage{amsmath,amsfonts} 
\usepackage{url}  
\usepackage{graphicx}
\usepackage{color}
\usepackage{subfig}
\usepackage{algorithm} 
\usepackage{algpseudocode}
\usepackage{rotating}
\usepackage{multirow}
\usepackage{booktabs}
\usepackage[round]{natbib}
\usepackage{authblk}

\newcommand{\col}{\text{col}}

\newcommand{\tr}{\text{tr}}

\newcommand{\rank}{\text{rank}}
\newcommand{\diag}{\text{diag}}

\newcommand{\blkdiag}{\text{blkdiag}}
\newcommand{\Ex}{\mathbb{E}} 
\newcommand{\real}{\mathbb{R}} 
\newcommand{\norm}{\mathcal{N}} 

\DeclareMathOperator*{\argmin}{arg\,min}

\def\bb{\boldsymbol{b}}

\def\bdf{\boldsymbol{f}}

\def\br{\boldsymbol{r}}

\def\bu{\boldsymbol{u}}
\def\bv{\boldsymbol{v}}

\def\bx{\boldsymbol{x}}
\def\by{\boldsymbol{y}}

\def\bA{\boldsymbol{A}}
\def\bB{\boldsymbol{B}}
\def\bC{\boldsymbol{C}}
\def\bD{\boldsymbol{D}}
\def\bE{\boldsymbol{E}}
\def\bF{\boldsymbol{F}}

\def\bI{\boldsymbol{I}}
\def\bJ{\boldsymbol{J}}

\def\bL{\boldsymbol{L}}

\def\bQ{\boldsymbol{Q}}
\def\bR{\boldsymbol{R}}
\def\bS{\boldsymbol{S}}

\def\bU{\boldsymbol{U}}
\def\bV{\boldsymbol{V}}
\def\bW{\boldsymbol{W}}
\def\bX{\boldsymbol{X}}
\def\bY{\boldsymbol{Y}}

\newcommand{\0}{\mathbf{0}}

\newcommand{\bbeta}{\boldsymbol{\beta}}

\newcommand{\bmu}{\boldsymbol{\mu}}

\newcommand{\bDelta}{\boldsymbol{\Delta}}

\newcommand{\bSigma}{\boldsymbol{\Sigma}}

\newcommand{\be}{\begin{equation}}
\newcommand{\ee}{\end{equation}}
\newcommand{\bea}{\begin{eqnarray}}
\newcommand{\eea}{\end{eqnarray}}
\newcommand{\bes}{\begin{eqnarray*}}
\newcommand{\ees}{\end{eqnarray*}}
\newcommand{\bi}{\begin{itemize}}
\newcommand{\ei}{\end{itemize}}

\newtheorem{thm}{Theorem}[section]

\newtheorem{prop}[thm]{Proposition}

\title{Incorporating Covariates into Integrated Factor Analysis of Multi-View Data}
\author[1]{Gen Li}
\author[2]{Sungkyu Jung}

\affil[1]{Department of Biostatistics, Mailman School of Public Health, Columbia University}
\affil[2]{Department of Statistics, University of Pittsburgh}
\renewcommand\Authands{ and }

\begin{document}

\date{}

\maketitle

\begin{abstract}
In modern biomedical research, it is ubiquitous to have multiple data sets measured on the same set of samples from different views (i.e., multi-view data). For example, in genetic studies, multiple genomic data sets at different molecular levels or from different cell types are measured for a common set of individuals to investigate genetic regulation. Integration and reduction of multi-view data have the potential to leverage information in different data sets, and to reduce the magnitude and complexity of data for further statistical analysis and interpretation. In this paper, we develop a novel statistical model, called supervised integrated factor analysis (SIFA), for integrative dimension reduction of multi-view data \textcolor{black}{while incorporating auxiliary covariates}. The model decomposes data into joint and individual factors, capturing the joint variation across multiple data sets and the individual variation specific to each set respectively. Moreover, both joint and individual factors \textcolor{black}{are partially informed by auxiliary covariates via nonparametric models}. We devise a computationally efficient Expectation-Maximization (EM) algorithm to fit the model under some identifiability conditions.  We apply the method to the Genotype-Tissue Expression (GTEx) data, and provide new insights into the variation decomposition of gene expression in multiple tissues. Extensive simulation studies \textcolor{black}{and an additional application to a pediatric growth study} demonstrate the advantage of the proposed method over competing methods. 
\end{abstract}


\section{Introduction}
In contemporary biomedical studies, researchers usually have access to multiple data sets for the same set of subjects from different views or heterogeneous sources.
Such data are commonly referred to as multi-view data or multi-source data.
For example, the Genotype-Tissue Expression (GTEx) project collects gene expression data from multiple human tissues for a common set of genotyped individuals to study genetic regulation \citep{ardlie2015genotype}.
Different data sets may contain distinct but related information.
It is important to understand the relations between variables in different sets, and leverage information across views for further statistical analysis such as inference, prediction and clustering.
The process is often called data integration or data fusion.

Factor analysis is a popular tool for modeling dependence among multiple observed variables.
It identifies a few latent factors that capture the majority of variation in data.
The unknown factors and loadings in factor analysis are sometimes estimated via the principal component analysis (PCA).
The obtained factors reduce the dimensionality of the original data and facilitate various statistical analyses.
However, the conventional factor analysis only applies to a single data set.
There is a pressing need for statistical methods that simultaneously identify the joint and individual structure in multiple data sets.

In addition to multiple primary data sets, auxiliary covariates are often collected on the same samples.
In our motivating GTEx example, other than the gene expression data in multiple tissues, genotype data and experimental factors (e.g., batch effect) are also collected.
These auxiliary data can be viewed as covariates, driving the underlying expression patterns in multiple tissues.
Covariates are potential driving factors of the joint and individual structures in multi-view data.
In other words, covariates provide {\em supervision} to the underlying patterns.
Using covariates to inform the integration of multi-view data not only leads to accurate estimation of the underlying patterns but also provides highly interpretable results.

In this paper, we develop a novel statistical framework called {\em Supervised Integrated Factor Analysis} (SIFA), for the integration and reduction of multi-view data informed by auxiliary covariates.
SIFA decomposes multi-view data into low-rank joint structure and individual structure.
It exploits a small number of joint factors to capture the shared patterns across all data sets, and separate individual factors to capture the specific patterns in each data set.
Corresponding loading vectors identify the contribution of the variables to different factors.
\textcolor{black}{To allow auxiliary covariates to inform the latent structure, the model assumes each factor is potentially driven by the covariates and some random effects.
We particularly consider regression models that flexibly accommodate parametric or nonparametric relations between factors and covariates.
Through the regression models, the covariates exert supervision on the latent structure.}
We also extend the model to incorporate variable selection, in order to identify important covariates that drive different factors.
Overall, SIFA provides a general framework for the covariate-driven factor analysis  of multi-view data.


There is an extensive body of literature on the integrative analysis of multi-view data \citep{tseng2015integrating}.
\textcolor{black}{Here we particularly focus on data integration and reduction.
Multiple factor analysis is an extension of the conventional factor analysis to multiple data sets \citep{abdi2013multiple}.
The idea is to merge multiple data with weights and perform the factor analysis on the combined data. However, the method does not distinguish joint and individual structure and may lead to misleading results.
More recently, new methods have been developed to decompose the total variation of multiple data sets into shared and individual variation  \citep{lofstedt2013global,ray2014bayesian,schouteden2014performing,yang2015non,zhou2015group}.
For example, \cite{lock2013joint} adopts an iterative PCA approach to estimate the Joint and Individual Variation Explained (JIVE).
However, a drawback of these methods is that they cannot take into account any auxiliary covariates in dimension reduction.
When covariates are strongly associated with the latent structure of the multi-view data, incorporating the supervision effects from covariates promises to improve estimation accuracy and interpretability.
}

Recently, a couple of methods were proposed to allow covariates to inform factor analysis. \cite{li2016superviseda} developed the Supervised Singular Value Decomposition (SupSVD) method that exploits linear models to accommodate covariates in dimension reduction of a primary data matrix.
Later, \cite{fan2016projected} proposed the projected PCA that generalizes SupSVD by allowing nonparametric relations between covariates and factors.
However, these methods are only suitable for a single  data set, and cannot easily extend to multi-view data.
To our best knowledge, there is no covariate-driven factor analysis method for multi-view data decomposition.
Our proposed method will bridge the gap and provide a unified framework.

The rest of the paper is organized as follows. In Section \ref{sec:model}, we propose a semiparametric latent variable model for SIFA and develop an Expectation-Maximization (EM) algorithm to fit the model.
In Section \ref{sec:sim}, extensive simulation studies are conducted to compare the proposed method with existing  methods.
In Section \ref{sec:real}, we apply SIFA to the GTEx multi-tissue genetic data to offer novel insights into the decomposition of genetic variation in a gene set across multiple tissues.
In Section \ref{sec:discuss}, we discuss possible directions for future research.
Technical details, additional simulation results, and an application to the decoupled growth amplitude and phase data from the Berkeley Growth Study can be found in the online supplementary material.

\section{\textcolor{black}{Integrated Factor Analysis Framework}}\label{sec:model}
In this section, we first introduce the latent variable model for SIFA and discuss its connection to existing methods. Then we elaborate two sets of identifiability conditions, and devise model fitting algorithms under respective conditions. Finally we propose rank selection methods to determine the joint and individual ranks in the model.

\subsection{Model}

Let $\bY_1,\cdots,\bY_K$ be $K$ primary data matrices of size $n\times p_1,\cdots, n\times p_K$ for the same set of samples collected from $K$ different sources.
Each row corresponds to a sample and each column is a variable.
Let $\bX$ be an $n\times q$ data matrix containing covariates for the matched samples.
The goal is to identify low-rank joint and individual patterns from the primary data matrices while accounting for the supervision effects from the covariates.
Without loss of generality, we center each column of the primary data and the covariates to remove the mean effect of each variable.


We propose a latent variable model called SIFA for the integrative factor analysis of multiple data matrices.
For $k=1,\cdots,K$, the SIFA model is as follows (without special notice, the index $k$ takes integer values from 1 to $K$):
\begin{eqnarray}\label{eq1}
\bY_k&=&\bJ_k+\bA_k+\bE_k,\\
\label{eq2}
\bJ_k&=&\bU_0\bV_{0,k}^T,\\
\label{eq3}
\bA_k&=&\bU_k\bV_k^T,\\
\label{eq4}
\bU_0&=&\bdf_0(\bX)+\bF_0,\\
\label{eq5}
\bU_k&=&\bdf_k(\bX)+\bF_k.
\end{eqnarray}

In \eqref{eq1}, we adopt a signal-plus-noise model to capture the important patterns in each data set.
This type of model is commonly used in the dimension reduction literature \citep[cf.][]{Shabalin2013}.
More specifically, the data matrix $\bY_k$ consists of signal $\bJ_k+\bA_k$ and noise $\bE_k$.
The matrix $\bJ_k$ captures the joint structure shared across multiple sources, and the matrix $\bA_k$ captures the individual structure specific to this data source.
The noise matrix $\bE_k$ is assumed to have independent and identically distributed (i.i.d.) entries from a normal distribution $\mathcal{N}(0,\sigma_k^2)$, where $\sigma_k^2$ is an unknown parameter.

In \eqref{eq2} and \eqref{eq3}, we assume that the joint and individual patterns for $\bY_k$ have low-rank decomposition.
Let $r_0$ be the underlying rank of the joint structure and $r_k$ be the rank of the individual structure in the $k$th data set.
Correspondingly, $\bU_0$ and $\bU_k$ are $n\times r_0$ and $n\times r_k$ (latent) factor matrices, and $\bV_{0,k}$ and $\bV_k$ are $p_k\times r_0$ and $p_k\times r_k$ loading matrices.
In particular, $\bU_0$ contains $r_0$ joint factors shared across different data sets, and $\bV_0=(\bV_{0,1}^T,\cdots,\bV_{0,K}^T)^T$ contains $r_0$ corresponding joint loadings.
The matrices $\bU_k$ and $\bV_k$ contain $r_k$ individual factors and loadings respectively.
Following the convention of the factor analysis, we assume the factors are independent and the loadings are orthonormal within each matrix.
Namely, $\bV^T_k\bV_k=\bI_{r_k}$ for each $k=0,1,\cdots,K$, where $\bI_{r_k}$ denotes the $r_k\times r_k$ identity matrix (we shall drop the subscript when it does not cause any confusion).

In order to capture the driving effects of covariates on the low-rank structure, we propose to  regress the latent factors on the covariates via \eqref{eq4} and \eqref{eq5}.
The mapping functions $\bdf_k(\cdot): \real^q\mapsto\real^{r_k}$ ($k=0,1,\cdots,K$) are unknown parametric or nonparametric functions.
With a slight abuse of notation, we use $\bdf_k(\bX)$ ($k=0,1,\cdots,K$) to represent row-wise mappings.
Namely, $\bdf_k(\bX)$ is an $n\times r_k$ matrix whose $i$th row corresponds to $\bdf_k(\bx_{(i)})$, where $\bx_{(i)}$ is the $i$th row of $\bX$ ($i=1,\cdots,n$).
The mapping functions capture flexible relations between covariates and the latent factors.
In practice, users can determine whether to use nonparametric functions or parametric functions (e.g., linear functions).
Any unknown variation in the factors is contained in the random matrices $\bF_k$ ($k=0,1,\cdots,K$).
In particular, we assume each row of $\bF_k$ follows an i.i.d.\ multivariate normal distribution with zero mean and an unknown diagonal covariance matrix $\bSigma_k$
(with positive, distinct, and decreasing diagonal values).
Moreover, we assume $\bF_0$, $\bF_k$'s, and $\bE_k$'s are mutually independent.

The proposed SIFA model provides a general framework for the factor analysis of multi-view data.
After rearranging the formulas, we get an equivalent form of the model as
\begin{equation}\label{model:SIPCA}
\bY_k=\bdf_0(\bX)\bV_{0,k}^T+\bdf_k(\bX)\bV_k^T +\bF_0\bV_{0,k}^T+\bF_k\bV_k^T+\bE_k.
\end{equation}
It is easy to see that the SIFA model decomposes the $k$th data matrix $\bY_k$ into five parts:
1) $\bdf_0(\bX)\bV_{0,k}^T$ is the joint deterministic structure (because $\bdf_0(\bX)$ is shared across multiple data sources and non-random) driven by the covariates;
2) $\bdf_k(\bX)\bV_k^T$ is the individual deterministic structure;
3) $\bF_0\bV_{0,k}^T$ is the joint random structure capturing any unknown variation; 
4) $\bF_k\bV_k^T$ is the individual random structure;
5) $\bE_k$ is the random noise.
With proper identifiability conditions which we will discuss later, the SIFA model attributes the total variation to different parts. Different model components will facilitate different analyses.
For example, the joint factors in $\bdf_0(\bX)+\bF_0$ can be potentially used for consensus clustering; 
the individual loadings in $\bV_k$ can be used to investigate the dependence among variables in the $k$th data source.

We remark that the proposed SIFA model \eqref{model:SIPCA} subsumes many existing methods as special cases.
When $K=1$, i.e., with only one primary data set $\bY$, there is no distinction between the joint structure and the individual structure.
Consequently, the model degenerates to
\[
\bY=(\bdf(\bX)+\bF)\bV^T+\bE,
\]
which corresponds to the projected PCA model proposed by \cite{fan2016projected}.
In particular, if we let the function $\bdf(\cdot)$ be a linear mapping, i.e., $\bdf(\bX)=\bX\bB$, where $\bB$ is a $q\times r$ coefficient matrix, the above model further connects to the SupSVD model developed in \cite{li2016superviseda}.
Furthermore, if we eliminate the covariate supervision by setting $\bdf(\bX)=\0$, the model degenerates to the conventional factor analysis model or  the probabilistic PCA model \citep{tipping1999probabilistic}.
When $K\geq2$, without accounting for the covariates (i.e., $\bdf_k(\bX)=\0;k=0,1,\cdots, K$), the SIFA model reduces to
\[
\bY_k=\bF_0\bV^T_{0,k}+\bF_k\bV_k^T+\bE_k.
\]
This coincides with the JIVE model \citep{lock2013joint} if we assume $\bF_0$ and $\bF_k$ are unknown score matrices for the joint and individual structures.
The SIFA model unifies and generalizes the above models, and provides a general framework for the integration and reduction of multi-view data informed by covariates.

\subsection{Identifiability}

Suppose $\theta_0=\{\bdf_0(\cdot),\bdf_k(\cdot),\bV_{0},\bV_{k},\bSigma_0,\bSigma_k,\sigma_k^2;k=1,\ldots,K\}$ is a parameter set for Model \eqref{model:SIPCA}, satisfying the basic conditions of $\bV_k^T\bV_k=\bI$ and $\bSigma_k$ being diagonal with distinct (decreasing) positive diagonal values for each $k=0,1,\cdots,K$.
If there is only one primary data set (i.e., $K=1$), the model is uniquely defined \citep{li2016superviseda}.
However, when there are multiple data sets, the above basic conditions are no longer sufficient for identifiability.

To be specific, let $\Theta$ be the collection of parameter sets $\theta$ satisfying the basic conditions and having equal likelihood $\mathcal{L} (\bY_1,\cdots,\bY_K\mid \theta)$ (defined later in \eqref{likfun}) with $\theta_0$ for any data $\bY_1,\cdots,\bY_K$.
Namely, any parameter set $\theta \in \Theta$ and $\theta_0$  are \emph{observationally equivalent} for Model \eqref{model:SIPCA}, i.e., $\Theta$ is the {\em equivalence class} of $\theta_0$.
We note that the collection $\Theta$ typically contains multiple elements (see the supplementary material for examples of some equivalent models).
In other words, $\theta_0$ is unidentifiable.
This non-identifiability is mainly caused by the indistinguishable individual and joint structures.
Different elements in $\Theta$ may have different sets of ranks, or the same set of ranks but different parameters.
Additional regularity conditions are needed to enforce the identifiability of the SIFA model.
For this, we propose two sets of sufficient conditions.

First, we consider a set of {\em general conditions} for each $k=1,\cdots,K$:
\bi
\item[A1.] Each submatrix $\bV_{0,k}$ of the joint loading matrix $\bV_0$ has full column rank;
\item[A2.] The columns in $\bV_{0,k}$ and $\bV_k$ are linearly independent, and $r_0+r_k<p_k$.
\ei
Loosely speaking, Condition A1 guarantees that the joint loading matrix $\bV_0$ indeed captures the joint structure, and does not contain any structure only pertaining to a subset of the $K$ data sets.
Condition A2 ensures that the joint and individual patterns are well separated, and are not interchangeable.
With both conditions, Model \eqref{model:SIPCA} is identifiable as shown in the following proposition (the proof is postponed to the supplementary material).

\begin{prop}\label{prop:id}
Let $\theta_0$ be a parameter set satisfying Conditions A1 and A2.
For any element $\theta$ in the equivalent class $\Theta$ of $\theta_0$, if $\theta$ also satisfies Conditions A1 and A2, then $\theta$ is equal to $\theta_0$ up to trivial sign changes.
Moreover, by writing $r_0(\theta)$ as the rank of $\bV_0$ in the parameter set $\theta$, we have $r_0(\theta_0)\leq r_0(\theta)$ for all $\theta\in\Theta$.
\end{prop}

The proposition guarantees that the general conditions are sufficient for model identifiability.
The identifiability is defined up to trivial column-wise sign changes in $\bV_k$ and $\bU_k$ ($k=0,1,\cdots,K$).
In practice, one could easily fix the signs by setting the first nonzero entry of each column of $\bV_k$ to be positive.
Correspondingly, the sign of each column of $\bU_k$ is fixed.

{\em Remark: Technically, the general conditions may slightly affect the generalizability of the model. Condition A1 rules out the possibility of any partially joint structure pertaining to multiple but not all data sets.
Namely, the model cannot identify common patterns across a subset of data sets.
The same issue exists for JIVE  as well.
This is a future research direction as discussed in Section \ref{sec:discuss}.
Nevertheless, in practice, the general conditions are suitable for most applications.}

In some circumstances, it is desired to further restrict the model parameters for better interpretation and computation.
We consider the following {\em orthogonal conditions}:
\bi
\item[B1.] The columns of $\bV_{0,k}$ are orthogonal with norm $1/\sqrt{K}$, i.e., $\bV_{0,k}^T\bV_{0,k}={1\over K}\bI$;
\item[B2.] The columns in $\bV_{0,k}$ and $\bV_k$ are orthogonal ($\bV_{0,k}^T\bV_k=\0$), and $r_0+r_k<p_k$.
\ei
Apparently, Conditions B1 and B2 are sufficient conditions for Conditions A1 and A2. Therefore, they are also sufficient conditions for the identifiability of the SIFA model.
Condition B1 implies that different data sets contribute roughly equally to the joint factors $\bU_0$ (i.e., columns in $\bV_{0,k_1}$ and $\bV_{0,k_2}$ have the same $\ell_2$ norm, for $k_1\neq k_2$).
In many real applications (e.g., the GTEx example in Section \ref{sec:real}), when the data are properly preprocessed, the equal contribution assumption can be well justified.
Conditions B1 and B2 also imply that the combined loadings $(\bV_{0,k},\bV_k)$ for the $k$th data set are mutually orthogonal.
For high dimensional data, it is reasonable to assume that the orthogonality between different loadings holds \citep{ahn2010maximal}.
When both assumptions are justified, it is beneficial to study the SIFA model under the orthogonal conditions.
These conditions not only improve model interpretation, but also facilitate computation as discussed in the next subsection.

\textcolor{black}{\em Remark: The SIFA model with the general conditions is equivariant under individual scaling of each data set. In other words, at the population level, the model is not affected by weighing multiple data sets differently. In practice, to avoid numerical instability, it is recommended to normalize different data sets to the same scale before estimation (e.g., set the Frobenius norm of every data set to be 1). The orthogonal conditions do not have the equivariant property under rescaling. Thus, if the orthogonal assumptions are justifiable, one should directly apply the method without scaling the data. See the supplementary material for more details. \textcolor{black}{There, we also discuss the effect of imbalanced dimensions of different data sets.}}

\subsection{Algorithm}

To estimate the model parameters in $\theta_0=\{\bdf_0(\cdot),\bdf_k(\cdot),\bV_{0,k},\bV_{k},\bSigma_0,\bSigma_k,\sigma_k^2; k=1,\ldots,K\}$ for Model \eqref{model:SIPCA}, we use a maximum likelihood approach.
We assume all random variables are from normal distributions.
For the ease of presentation, $\bV_\star=\mbox{blkdiag}(\bV_1,\cdots,\bV_K)$ denotes the combined individual loading matrix of size $\sum_{k=1}^K p_k \times \sum_{k=1}^K r_k$, which is a block-wise diagonal matrix with $K$ diagonal blocks $\bV_1,\cdots,\bV_K$.
We also let $\bU_\star=(\bU_1,\cdots,\bU_K)=(\bdf_1(\bX)+\bF_1,\cdots,\bdf_K(\bX)+\bF_K)$ denote the combined individual factor matrix.
Let $\bY_\star=(\bY_1,\cdots,\bY_K)$ and $\bE_\star=(\bE_1,\cdots,\bE_K)$ be the combined primary data matrix and noise matrix respectively.
As a result, the SIFA model can be succinctly expressed as
\begin{equation*}\label{model:simplified}
\bY_\star=\bU_0\bV_0^T +\bU_\star\bV_\star^T+\bE_\star.
\end{equation*}

The log likelihood function of the SIFA model is
\begin{eqnarray}\label{likfun}
\log\mathcal{L}( \bY_\star\mid\theta_0)=\sum_{i=1}^n\left[ -{\sum_{k=1}^K p_k\over2}\log 2\pi -{1\over2} \log|\bSigma_\star| - {1\over2}(\by_{\star(i)}-\bmu_{\star(i)})^T\bSigma_\star^{-1}(\by_{\star(i)}-\bmu_{\star(i)})\right],
\end{eqnarray}
where $\by_{\star(i)}$ is a column vector corresponding to the $i$th row of $\bY_\star$, and $\bmu_{\star(i)}$ and $\bSigma_\star$ are the marginal expectation and covariance matrix of $\by_{\star(i)}$ respectively.
In particular,
\[
\bmu_{\star(i)}^T=\bdf_0(\bx_{(i)})\bV_0^T+\left[\bdf_1(\bx_{(i)})\bV_1^T,\cdots,\bdf_K(\bx_{(i)})\bV_K^T\right],
\]
where $\bx_{(i)}$ is a column vector corresponding to the $i$th row of $\bX$, and $\bdf_k(\bx_{(i)})$ is a row vector of length $r_k$ ($k=0,1,\cdots,K$).
The grand covariance matrix $\bSigma_\star$ has the form
\[
\bSigma_\star=\bV_0\bSigma_0\bV_0^T+\bV_\star\bSigma_{\bF}\bV_\star^T+\bSigma_{\bE},
\]
where $\bSigma_{\bF}=\mbox{blkdiag}(\bSigma_1,\cdots,\bSigma_K)$ and $\bSigma_{\bE}=\mbox{blkdiag}(\sigma_1^2\bI_{r_1},\cdots,\sigma_k^2\bI_{r_K})$.
The optimization of the above log likelihood function under the identifiability conditions is computationally prohibitive because the likelihood function involves unknown nonparametric functions and the conditions are non-convex.

To circumvent the computational issue, we resort to the hierarchical form of the SIFA model in \eqref{eq1}--\eqref{eq5} and treat $\bU_0$ and $\bU_\star$ as latent variables, and derive an {\em Expectation-Maximization} (EM) algorithm.
Specifically, in the E step, we calculate the conditional distribution of the latent variables ($\bU_0,\bU_\star)$ given the data $\bY_\star$ and the previously estimated model parameters.
In the M step, we maximize the conditional expectation of the joint log likelihood of the latent variables and the data.
The joint log likelihood is partitioned into the log likelihood of $(\bU_0, \bU_\star)$ and the conditional log likelihood of $\bY_\star$ given $(\bU_0, \bU_\star)$.
Furthermore, since the latent variables $\bU_0,\bU_1,\cdots,\bU_K$ are mutually independent, the log likelihood of $(\bU_0,\bU_\star)$ is further partitioned.
Consequently, the M step is  to  solve the following problems under the respective identifiability conditions:
\begin{eqnarray}\label{opt:M1}
&\max\limits_{\bdf_k(\cdot),\bSigma_k} \ &\ \mathbb{E}_{\bU_k|\bY_\star} \, \mathcal{L}(\bU_k), \quad k=0,1,\cdots,K,\\
\label{opt:M2}
&\max\limits_{\bV_0,\bV_\star,\sigma_1^2,\cdots,\sigma_K^2} \ &\ \mathbb{E}_{\bU_0,\bU_\star|\bY_\star} \, \mathcal{L}(\bY_\star|\bU_0,\bU_\star),
\end{eqnarray}
where $\Ex_{\bU_0,\bU_\star|\bY_\star}(\cdot)$ represents the conditional expectation with respect to $(\bU_0,\bU_\star)$.
For simplicity, hereafter we will use $\mathbb{E}(\cdot)$ to denote the conditional expectations.
Below we shall outline the key steps of the M step. More details can be found in the supplementary material.

It can be shown that in \eqref{opt:M1} each entry of the vector-valued function $\bdf_k(\cdot)=(f_{k,1}(\cdot),\cdots,f_{k,r_k}(\cdot))$ can be separately estimated via solving a least square problem
\be\label{reg}
\widehat{f_{k,j}(\cdot)}=\argmin_{f_{k,j}(\cdot)}\ \|\mathbb{E}(\bu_{k,j})-f_{k,j}(\bX)\|_\mathbb{F}^2, \quad j=1,\cdots,r_k;\ k=0,1,\cdots,K,
\ee
where $\bu_{k,j}$ is the $j$th column of $\bU_k$, and $\|\cdot\|_\mathbb{F}$ denotes the Frobenius norm.
If $f_{k,j}(\cdot)$ is a parametric function, the above problem can be solved via a Newton-Raphson method.
In particular, if linear, it is explicitly solved via the ordinary least squares.
If $f_{k,j}(\cdot)$ is nonparametric, the problem becomes nonparametric regression.
Standard kernel methods and spline-based methods can be readily applied here \citep[cf.][]{fan1996local,hollander2013nonparametric}.
When the dimension of the covariates is high, we can assume $f_{k,j}(\cdot)$ to be an additive model and easily incorporate variable selection through penalized methods \citep{tibshirani1996regression,ravikumar2009sparse}.
To sum up, regardless of the forms of the functions, $\{\bdf_k(\cdot)\}_{k=0,\cdots,K}$  can be easily estimated using existing methods.

Subsequently, it is easy to obtain the closed-form optimizer of \eqref{opt:M1} with respect to $\bSigma_k$ as:
\[
\widehat{\bSigma_k}={1\over n}\diag\left\{\mathbb{E}\left[\left(\bU_k-\widehat{\bdf_k}(\bX)\right)^T\left(\bU_k-\widehat{\bdf_k}(\bX)\right)\right]\right\},\ k=0,1,\cdots,K,
\]
where $\diag(\bS)$ is the diagonal matrix consisting of the diagonal values of $\bS$, and $\widehat{\bdf_k}(\cdot)$'s are the estimated covariate functions.

From \eqref{opt:M2}, we obtain the estimates of the loading matrices and the noise variances under different identifiability conditions.

Under the general conditions A1 and A2, there are no explicit solutions of \eqref{opt:M2} for $\bV_0$ and $\bV_\star$.
We propose to iteratively update the estimates of the loading matrices in a block-wise coordinate descent fashion.
In particular, we cycle through the following steps: given $\bV_0$, update $\bV_k$'s in parallel via solving
\begin{eqnarray}\label{opt:M3}
&\min\limits_{\bV_k:\bV_k^T\bV_k=\bI}&\ \mathbb{E}\|\bY_k-\bU_0\bV_{0,k}^T-\bU_k\bV_k^T\|_\mathbb{F}^2;
\end{eqnarray}
and given $\bV_\star$, update $\bV_0$ by solving
\begin{eqnarray}\label{opt:M4}
&\min\limits_{\bV_0:\bV_0^T\bV_0=\bI}&\ \sum_{k=1}^K \sigma_k^{-2}\mathbb{E}\|\bY_k-\bU_k\bV_{k}^T-\bU_0\bV_{0,k}^T\|_\mathbb{F}^2.
\end{eqnarray}
It can be shown that the optimization problem \eqref{opt:M3} has a closed-form solution
$
\widehat{\bV_k}=\bL\bR^T,
$
where $\bL$ and $\bR$ contain the left and right singular vectors of $\bY_k^T\mathbb{E}(\bU_k)-\bV_{0,k}\mathbb{E}(\bU_0^T\bU_k)$.
The optimization \eqref{opt:M4} does not have an analytical solution due to the possibly different  $\sigma_k^2$'s.
As a remedy, we relax the orthogonality constraint in \eqref{opt:M4} temporarily, and obtain an intermediate estimator of $\bV_{0,k}$ as
\[
\widetilde{\bV_{0,k}}=\left[\bY_k^T\mathbb{E}(\bU_0)-\bV_k\mathbb{E}(\bU_k^T\bU_0)\right]\left[\mathbb{E}(\bU_0^T\bU_0)\right]^{-1}.
\]
To impose the orthogonality constraint, the final estimator of $\bV_0$ is the eigenvectors of $\widetilde{\bV_0}\widehat{\bSigma_0}{\widetilde{\bV_0}}^T$.
Correspondingly, we update the diagonal values of $\widehat{\bSigma_0}$ to be the eigenvalues of $\widetilde{\bV_0}\widehat{\bSigma_0}{\widetilde{\bV_0}}^T$.
This additional standardization step ensures that $\bSigma_\star$ in the likelihood function \eqref{likfun} remains unchanged.
A similar approach was used in \cite{li2016superviseda}.
As a result, the loading matrices are estimated under the general conditions.
We remark that in practice, a one-step update in each EM iteration is usually accurate enough and there is no need to iterate.

Under the orthogonal conditions B1 and B2, the computation can be greatly simplified.
The loading matrices $\bV_{0,k}$ and $\bV_k$ can be estimated simultaneously and explicitly.
By writing $\bW_k=(\sqrt{K}\bV_{0,k},\bV_k)$, the optimization \eqref{opt:M2} is equivalent to
\[
\min_{\bW_k:\bW_k^T\bW_k=\bI} \left\|\bY_k-\left({1\over\sqrt{K}}\mathbb{E}(\bU_0),\mathbb{E}(\bU_k)\right)\bW_k^T\right\|_\mathbb{F}^2,
\]
which is exactly an orthogonal Procrustes problem \citep{gower2004procrustes}.
The optimizer has the explicit expression as $\widehat{\bW_k}=(\sqrt{K}\widehat{\bV_{0,k}},\widehat{\bV_k})=\bL\bR^T$, where $\bL$ and $\bR$ contain the left and right singular vectors of $\bY_k^T\left({1/\sqrt{K}}\mathbb{E}(\bU_0),\mathbb{E}(\bU_k)\right)$.
Subsequently, it is easy to decouple $\widehat{\bV_{0,k}}$ and $\widehat{\bV_k}$, and obtain closed-form estimators for different loading matrices.

Once the loading matrices are estimated, solving \eqref{opt:M2} with respect to $\sigma_k^2$'s, we obtain the closed-form optimizers as:
\[
\widehat{\sigma_k^2}={1\over np_k} \mathbb{E}\|\bY_k-\bU_k\widehat{\bV_{k}}^T-\bU_0\widehat{\bV_{0,k}}^T\|_\mathbb{F}^2, \ k=1,\cdots,K.
\]

\textcolor{black}{A step-by-step description of the algorithm can be found in the supplementary material.}

\subsection{Rank Selection}\label{rank}
Up to now, we assume the ranks for the joint and individual structures in the SIFA model are known.
In practice, we often need to estimate the ranks from data.
The choice of the ranks is crucial for parameter estimation and model interpretation.
In total, there are $K+1$ ranks to be determined.
\textcolor{black}{Here we propose a two-step procedure to get a crude estimate of the ranks, and an optional likelihood cross validation (LCV) method for refining the estimate. }

Since Model \eqref{model:SIPCA} can be viewed as a special form of a signal-plus-noise model, a natural first step is to estimate the rank of the underlying signal of each data set $\bY_k$ (denoted as $r_k^\star$) and the rank of the underlying signal of the combined data set $\bY_\star$ (denoted as $r_{total}^\star$) respectively.
\textcolor{black}{There are many existing methods to this purpose, such as the scree plot, the total variance explained criterion, hypothesis testing methods. 
Users can choose their favorite methods. }
Once estimated, we use $r_k^\star$ and $r_{total}^\star$ to calculate the ranks for different structures in Model \eqref{model:SIPCA}.
Under either set of identifiability conditions, the following equations hold:
\begin{equation*}
r_{total}^\star=r_0+\sum_{k=1}^K r_k,\quad r_k^\star= r_0+r_k,
\end{equation*}
for $k=1,\cdots,K$, where $r_0,r_1,\cdots,r_K$ are the joint and individual ranks for the SIFA model.
%
Solving the equation system, we get
\[
r_0= {\sum_{k=1}^K r_k^\star-r_{total}^\star\over K-1},\quad r_k= r_k^\star-r_0,
\]
which serve as good initial estimators of the ranks.
Numerically, the estimate of $r_0$ may be non-integer or even negative when $K>2$.
In that case, we suggest rounding the estimate to the nearest non-negative integer.
Then we plug it into the second equation to get an estimate of $r_k$.
If the estimate of $r_k$ is negative, it can be set to $0$.

\textcolor{black}{The above two-step procedure typically provides a good initial estimate of the ranks. If it is desired to further refine the rank estimation, one may exploit a more computationally intensive $N$-fold LCV approach. The idea is to randomly split the samples into $N$ groups across different data sets. In each run, we withhold one group as the testing set and use the remaining $N-1$ groups as the training set to fit Model \eqref{model:SIPCA} with different sets of ranks. \textcolor{black}{For each set of ranks, the corresponding LCV score is the value of negative log likelihood, evaluated using \eqref{likfun} on the testing data. We repeat the procedure $N$ times, and choose the set of ranks corresponding to the smallest average LCV score. A more detailed description can be found  in the supplementary material.} }

\section{Simulation Studies}\label{sec:sim}
In this section, we conduct comprehensive simulation studies to demonstrate the advantage of the proposed methods. \textcolor{black}{We compare SIFA (under both sets of identifiability conditions) with JIVE (the original version and a covariate-adjusted version, denoted by cov-JIVE), SupSVD, and PCA.
For cov-JIVE, we first regress different data sets on the covariates, and then apply JIVE to the residuals. }

\subsection{Simulation Settings}
We consider two primary data sets $\bY_1$ and $\bY_2$ (i.e., $K=2$) on the same set of samples with sample size $n=500$, and dimension $p_1=p_2=200$. 
The data are simulated from Model \eqref{model:SIPCA} with different parameters.
We first consider 3 settings where, loosely speaking, the generative models are JIVE, SIFA under the general conditions (denoted as SIFA-A), and SIFA under the orthogonal conditions (denoted as SIFA-B).
In particular, the SIFA-A and SIFA-B models employ linear models between covariates (with dimension $q=10$) and latent factors.
The true ranks of the joint and individual patterns are $r_0=2$, $r_1=r_2=3$.
Some important features of these settings are described below.

\bi
\item {\bf Setting 1} (JIVE Model): For $k=0,1,2$, the factors in $\bU_k$ are randomly generated and mutually independent (with $\bdf_k(\cdot)=\0$ in \eqref{model:SIPCA}); the loadings in $\bV_k$ and the covariance $\bSigma_k$ satisfy the basic conditions.
    The measurement errors in $\bE_1$ and $\bE_2$ are i.i.d.\ with different variances.
    To test the robustness of the proposed method, we randomly generate 10 covariates  unrelated with the factors, and incorporate them in the SIFA estimation.
\item {\bf Setting 2} (SIFA-A Model): The joint and individual factors are generated from the linear model $\bU_k=\bX\bB_k+\bF_k$ for $k=0,1,2$.
    The loadings in $\bV_0,\bV_1$ and $\bV_2$ are filled with random numbers and standardized to satisfy the general conditions. To make them further deviate from the orthogonal conditions, we intentionally choose $\bV_{0,k}$ not  orthogonal to $\bV_{k}$ $(k=1,2)$, and artificially vary the norm of each column in $\bV_{0,1}$ and $\bV_{0,2}$.
\item {\bf Setting 3} (SIFA-B Model): The factors are generated in the same way as in Setting 2. The true loadings are specifically normalized to satisfy the orthogonal conditions. We note that the SIFA-B model is a special case of the SIFA-A model.
\ei


For each simulation setting, we fit JIVE, cov-JIVE, SIFA-A, and SIFA-B to the multiple simulated data sets, and fit PCA and SupSVD to the concatenated data $(\bY_1,\bY_2)$. We incorporate covariates for cov-JIVE, SIFA-A, SIFA-B, and SupSVD.
In particular, when fitting the SIFA models,  we set the functions in \eqref{reg} to be linear, and use the ordinary least squares to estimate the coefficients.
\textcolor{black}{To avoid ambiguity, these model models are fitted with the true ranks.
We set the rank for PCA and SupSVD to be $r_0+r_1+r_2$.
We assess the performance of the LCV for rank selection in the next section.}

To compare the loading estimation in JIVE, cov-JIVE, SIFA-A and SIFA-B, we use the Grassmannian metric \citep{mattila1999geometry}  between the true loadings in $\bV_k$ and the estimated loadings in $\widehat{\bV}_k$ for each $k=0,1,2$.
The metric is defined as $d_{\mathcal{G}}(\bV_k,\widehat{\bV}_k)=\sqrt{\sum_{i=1}^{r_k}\mbox{acos}(\delta_i)^2}$,
where $\delta_i$ is the $i$th singular value of ${\bV_k^T\widehat{\bV}_k}$.
We also evaluate the maximal principal angle $\angle(\bV,\widehat{\bV})$ \citep{bjorck1973numerical} between the subspaces in $\real^{p_1+p_2}$ spanned by the true  loading vectors in $\bV=(\bV_0,\mbox{blkdiag}(\bV_1,\bV_2))$ and the estimated ones, across all methods.
To evaluate the accuracy of the estimated low-rank structure, we use
$\|\bU\bV^T-\widehat{\bU}\widehat{\bV}^T\|_{\mathbb{F}}$ where $\bU=(\bU_0,\bU_1,\bU_2)$.
The matrix $\widehat{\bU}$ represents the estimated score matrix for PCA and JIVE, or the conditional expectation of the latent factor matrix  for SupSVD, SIFA-A, and SIFA-B.

\textcolor{black}{We also conduct comprehensive simulation studies to investigate: 1) the goodness of fit when the relations between covariates and latent factors are nonlinear; 2) the overfitting issue when nonparametric functions are used in the presence of linear relations;  3) the rank misspecification effect on the performance; 4) the violation of the Gaussian assumption; 5) the effect of rescaling different data sets; 6) the scalability of SIFA-A and SIFA-B in high dimension. The simulation settings and results are contained in the supplementary material.}

{\color{black}
\subsection{Rank Estimation by LCV}
We briefly demonstrate the efficacy of the LCV method using a simulated example.
Data are generated under Setting 3, with the chosen true ranks to be $(r_0,r_1,r_2)=(2,3,3)$.
Additional examples under Setting 1 and Setting 2 are provided in the supplementary material.
We particularly consider 9 candidate rank sets in the neighborhood of the true rank set: $(r_0,r_1,r_2)\in\{(1,2,2),(2,2,2),(3,2,2),(1,3,3),(2,3,3),\\(3,3,3),(3,4,3),(3,4,4),(4,4,4)\}$.
We conduct a 10-fold LCV.
The evaluated LCV scores (i.e., the negative log likelihood values of test samples) for different candidate sets in each cross validation run are shown in Figure \ref{fig:LCV}.
The average score reaches the minimum at the true rank set.
Namely, the LCV method correctly selects the true ranks.
\begin{figure}[htbp]
\begin{center}
\includegraphics[width=4in]{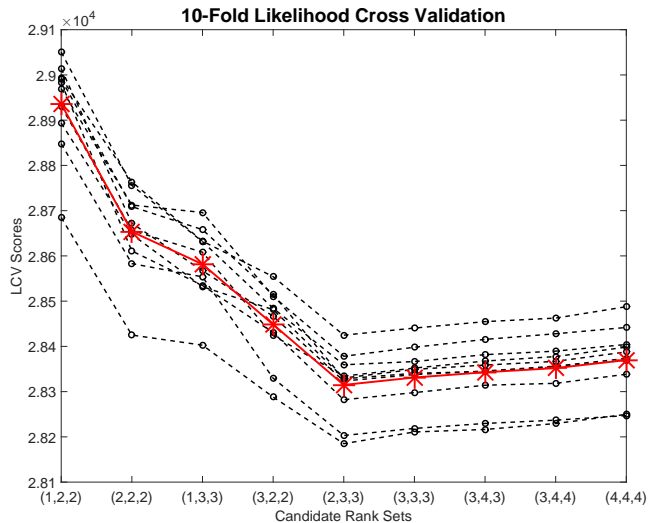}
\end{center}
\caption{The LCV scores for 10-fold cross validation on 9 candidate rank sets. Each dashed line with circles contains corresponds to the LCV scores (negative log likelihood values) in one cross validation run. The solid line with stars contains the average LCV scores for different rank sets. }
\label{fig:LCV}
\end{figure}
}

\subsection{Simulation Results}
For each setting, we repeat the simulation 100 times and summarize the results.
The results  are summarized in Table \ref{tab:sim}.
In Setting 1, both SIFA-A and SIFA-B perform similarly to JIVE in terms of the loading estimation, even if the generative model is JIVE (i.e., the covariates are unrelated to the factors).
Remarkably, the SIFA methods provide the best low-rank structure recovery accuracy among all.
The reason is similar to the argument in \cite{li2016superviseda}: the shrinkage effect of the conditional expectation of the factors reduces estimation variance.
In Setting 2, SIFA-A provides the best performance in all criteria.
SIFA-B is suboptimal because the orthogonal conditions are severely violated.
In Setting 3, the SIFA-B method performs the best, followed closely by SIFA-A.
Both are significantly better than the competing methods.
In practice, when the orthogonal conditions are well justified, SIFA-B is favorable due to the ultra-fast computation and accurate estimation.
Otherwise, we recommend the use of the SIFA-A method.

\begin{sidewaystable}[!]
\caption{Simulation results under Setting 1, 2 and 3 (each with 100 simulation runs). The mean and standard deviation of each criterion for each method are shown in the table. The best results are highlighted in bold.}
\label{tab:sim}
\begin{tabular}{l|c|c|c|c|c|c|c|c}
 & &  SIFA-A & SIFA-B & JIVE & cov-JIVE & SupSVD & PCA \\
\toprule
 \multirow{5}{*}{Setting 1}
       & $d_{\mathcal{G}}(\bV_0,\widehat{\bV}_0)$    &{ 0.61}(0.03)  &  {\bf 0.60}(0.03)  &  0.69(0.07) & 0.68(0.06)&& \\
 \multirow{5}{*}{(JIVE)}
      & $d_{\mathcal{G}}(\bV_1,\widehat{\bV}_1)$    &0.82(0.06)  &  {\bf 0.81}(0.06)  &  0.91(0.17) & 0.89(0.16)&& \\
       & $d_{\mathcal{G}}(\bV_2,\widehat{\bV}_2)$    &1.32(0.17)  &  1.33(0.17)  &  { 1.31}(0.18) &{\bf 1.30}(0.17)&& \\
    \cmidrule{2-8}
       & $\angle(\bV,\widehat{\bV})$   & { 64.73}(10.98) &   65.33(10.96) &   65.16(11.31) &  {\bf 64.35}(10.98) &  87.02(2.74) & 86.76(2.77)\\
    \cmidrule{2-8}
       & $\|\bU\bV^T-\widehat{\bU}\widehat{\bV}^T\|_{\mathbb{F}}$ &   193.28(2.85) &  {\bf 193.26}(2.73) &  240.49(4.53) & 287.86(3.74) &239.21(2.82) &  279.95(3.05)\\

\midrule

 \multirow{5}{*}{Setting 2}
       & $d_{\mathcal{G}}(\bV_0,\widehat{\bV}_0)$    &    {\bf 0.37}(0.02)  &  1.01(0.05) &   0.76(0.02) & 1.40(0.14) && \\
 \multirow{5}{*}{(SIFA-A)}
      & $d_{\mathcal{G}}(\bV_1,\widehat{\bV}_1)$    &    {\bf 0.27}(0.01) &   1.06(0.08)  &  0.28(0.01) & 1.41(0.03)&& \\
       & $d_{\mathcal{G}}(\bV_2,\widehat{\bV}_2)$    &    {\bf 0.52}(0.02) &   0.67(0.03) & 0.62(0.04) & 1.80(0.09)&& \\
    \cmidrule{2-8}
       & $\angle(\bV,\widehat{\bV})$   &    {\bf 27.67}(1.42) &  44.46(2.42)& 33.34(2.27) &  88.30(1.43)&  39.40(2.30) &   46.97(4.28)\\
    \cmidrule{2-8}
       & $\|\bU\bV^T-\widehat{\bU}\widehat{\bV}^T\|_{\mathbb{F}}$ &     {\bf 169.21}(1.73) &  207.74(2.03)  & 207.22(2.33) & 296.14(2.11)& 200.77(2.30) &  235.44(3.00) \\

\midrule

 \multirow{5}{*}{Setting 3}
       & $d_{\mathcal{G}}(\bV_0,\widehat{\bV}_0)$    &0.38(0.03)   & {\bf 0.30}(0.01)  &  0.61(0.02)  & 0.65(0.06)&& \\
 \multirow{5}{*}{(SIFA-B)}
       & $d_{\mathcal{G}}(\bV_1,\widehat{\bV}_1)$    &    0.25(0.01) &   {\bf 0.24}(0.01)  &  0.25(0.01) & 1.16(0.21) && \\
       & $d_{\mathcal{G}}(\bV_2,\widehat{\bV}_2)$    & 0.35(0.01)  &  {\bf 0.34}(0.01)  &  0.36(0.01) & 1.77(0.07) && \\
    \cmidrule{2-8}
       & $\angle(\bV,\widehat{\bV})$   &    15.03(0.56) &  {\bf 14.97}(0.57)  & 15.86(0.73)  & 85.93(2.83) & 26.21(1.09)  & 27.71(1.30)\\
    \cmidrule{2-8}
       & $\|\bU\bV^T-\widehat{\bU}\widehat{\bV}^T\|_{\mathbb{F}}$ &     171.99(1.65) & {\bf 171.51}(1.66)  &   204.64(1.97) & 290.82(3.33)& 200.77(1.81) & 230.80(2.05)\\
\bottomrule
\end{tabular}
\end{sidewaystable}

\section{GTEx Data Analysis}\label{sec:real}
\textcolor{black}{In this section, we apply the proposed method to the multi-tissue genetic data from the GTEx project.
We use the phs000424.v6 data which are available at \url{http://www.gtexportal.org/} (registration required for data access).
Technical details of data preprocessing and rank estimation can be found in the supplementary material. }

The GTEx project collects gene expression data from multiple tissues (e.g., muscle, blood, skin) from the same set of subjects.
{\color{black} We use the SIFA method to identify cross-tissue and tissue-specific gene expression patterns, and quantify the heritability of phenotypes representing expressions of a group of genes.
Addressing the questions is integral to the fulfillment of the GTEx goal \citep{ardlie2015genotype}.
}

\textcolor{black}{We particularly focus on the p53 signaling pathway in three tissues, i.e., muscle, blood, and skin, for the illustration purpose.
The analysis can be easily generalized to other gene sets or tissues.
After proper preprocessing and normalization, we obtain 191 genes on 204 common samples in each tissue, denoted by $\bY_1,\bY_2,\bY_3$.
Each gene expression is standardized.
In addition, we have the auxiliary data of sex, genotyping platform index, and genetic variants for each sample as covariates.
To reduce the dimensionality of the genetic variants, we obtain the top 30 principal components, which capture the majority of variation in the genotype data. 
The covariates are denoted by $\bX$.
 }

\textcolor{black}{
We first estimate the ranks for the joint and individual patterns.
We use the two-step procedure described in Section \ref{rank}, and exploit a variance explained criterion in the first step (with a preset 90\% threshold).
The joint and individual ranks are estimated to be $r_0=26$, $r_1=24$, $r_2=5$, and $r_3=20$.
Note that the individual rank for blood  ($r_2=5$) is much smaller than that for muscle or skin.
From the viewpoint of the expression pattern richness, blood is very different from the other two tissues. This is generally concordant with the previous discoveries \citep{ardlie2015genotype}
}

\begin{figure}[htbp]
\begin{center}
\includegraphics[width=5in]{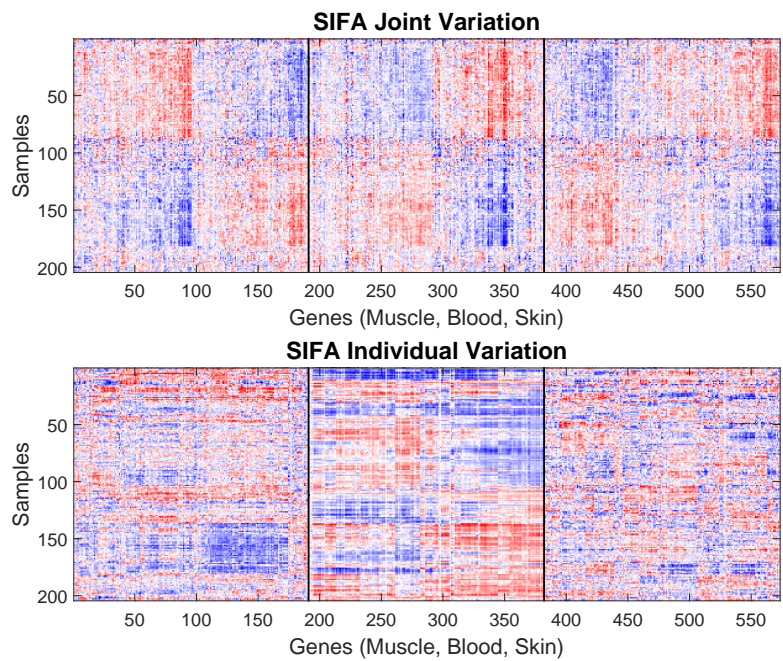}
\end{center}
\caption{GTEx Example: The heat maps of the joint and individual gene expression patterns for the p53 signaling pathway identified by the SIFA-B model. For visualization purpose, we reorder samples across three tissues and genes in each tissue.  Top panel: the joint structure in three tissues; Bottom panel: the individual structures in three tissues. In each panel, the samples are matched across tissues.}
\label{fig:decomp}
\end{figure}
\begin{figure}[htbp]
\begin{center}
\includegraphics[width=5in]{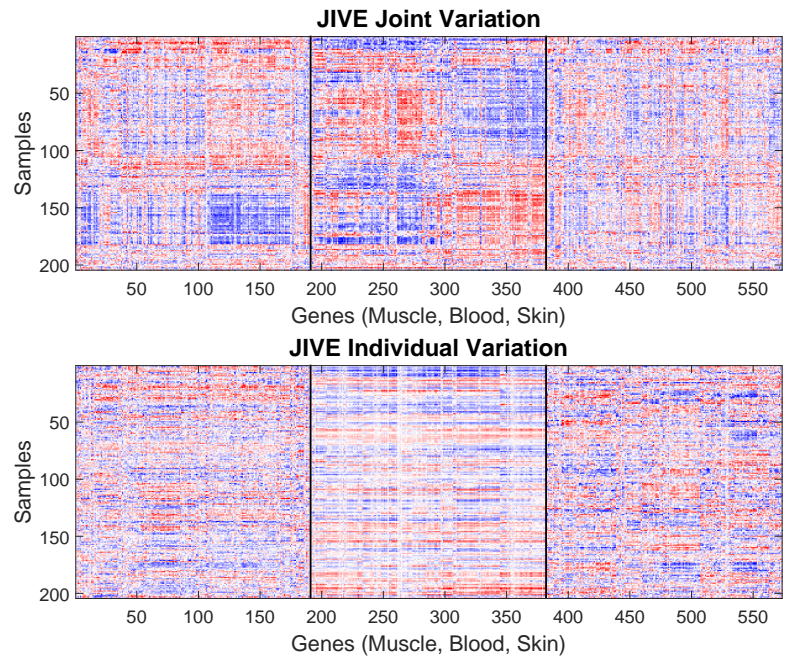}
\end{center}
\caption{GTEx Example: The heat maps of the joint and individual gene expression patterns for the p53 signaling pathway identified by the JIVE model. The rows and columns are ordered in the same way as in Figure \ref{fig:decomp}.}
\label{fig:JIVE}
\end{figure}
\begin{figure}[htbp]
\begin{center}
\includegraphics[width=5in]{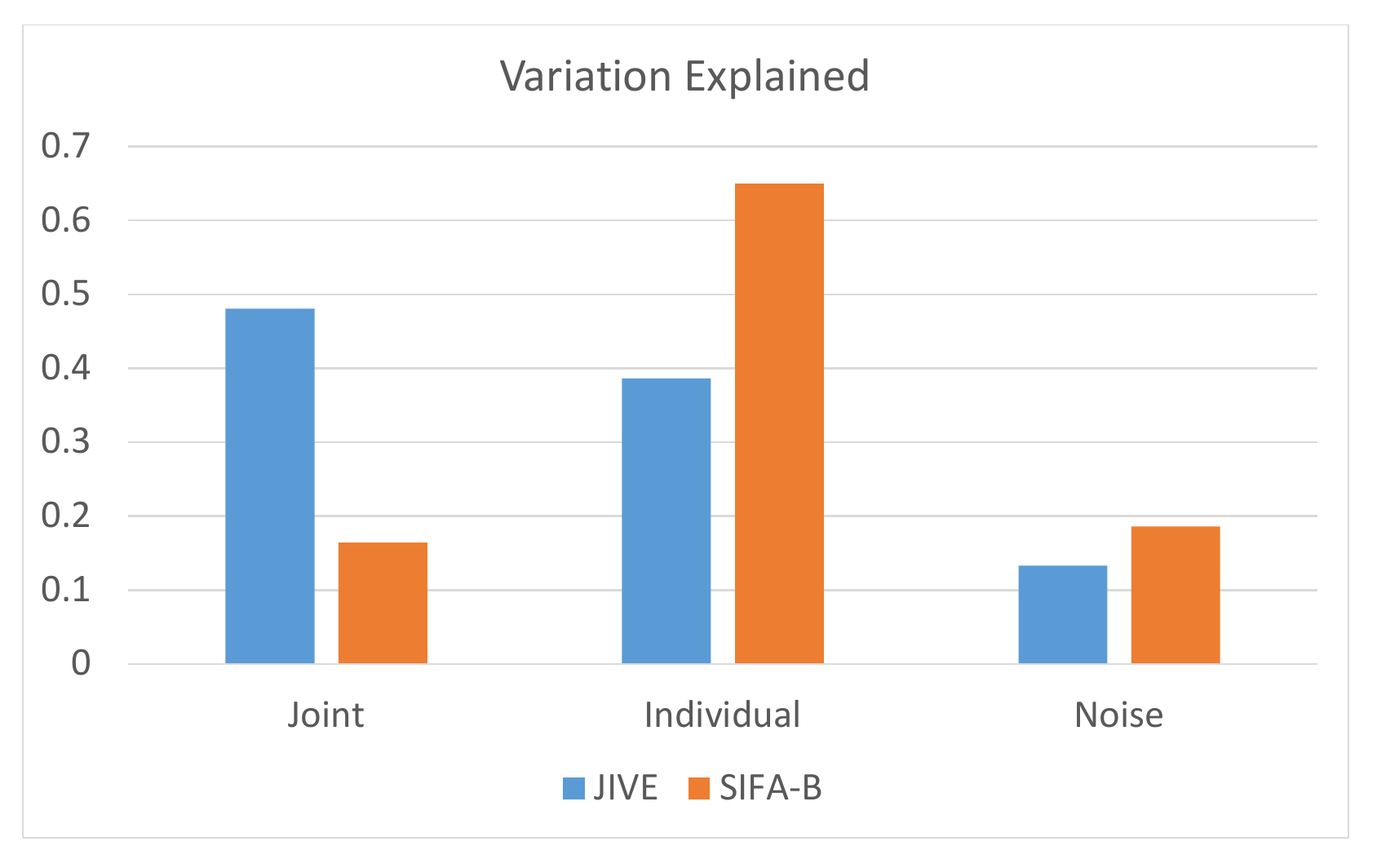}
\end{center}
\caption{GTEx Example: The variation explained by different components in the JIVE model and the SIFA-B model, respectively.}
\label{fig:var}
\end{figure}

\textcolor{black}{
We fit a SIFA-B model to the data with linear relations between the covariates and the latent factors.
For comparison, we also fit a JIVE model with the same ranks. 
The estimated joint and individual patterns are shown in Figure \ref{fig:decomp} (for the SIFA-B model) and Figure \ref{fig:JIVE} (for the JIVE model).
By taking into account the auxiliary covariates, the patterns obtained by the SIFA-B model are more discernable than those from the JIVE model.
The joint structure in Figure \ref{fig:decomp} clearly captures the shared patterns among samples across tissues, while the individual structure distinguishes different tissues.
We also quantify the variation explained by different parts in both methods (see Figure \ref{fig:var}).
The SIFA-B decomposition attributes more variation to the individual structure than the JIVE method, which is consistent with the domain knowledge that the p53 gene expressions are highly tissue specific \citep{ribeiro2001inherited,tendler1999tissue}.
The tissue-specific expression patterns may be used to investigate tissue identity and functions.}

\textcolor{black}{
To quantify the heritability of the derived phenotypes (i.e., joint and individual scores) representing the p53 gene expressions,
we calculate the variation explained by different components of the SIFA model.
The results are summarized in Table \ref{tab:GTEx}.
Within the joint structure (common across all tissues), the genetic variants explain about 17\% of the variation, which is concordant with the general belief in the literature \citep{brown2015pathway}.
The sex and the platform information take up $2\%$ and $2.5\%$ of the variation, respectively.
The vast majority of the variation remains unexplained, which provokes further investigation.
The individual structure for each tissue has a similar decomposition to the joint structure.
An interesting finding is that sex is not a major contributor to the individual gene expression patterns in blood.
The derived pathway expression phenotypes could also potentially be used
to discover associations with clinical outcome and environmental factors.
Due to the lack of such information in the GTEx data, we do not further pursue it here.
}

\begin{table}[h]
\caption{GTEx Example: The genetic variation explained by different factors in different tissues. For each tissue, the last column gives the percentage explained by the joint and individual structure, and the noise (add up to 1). The variation in the joint (individual) structure is further attributed to the genotype, sex, platform, and other unknown sources (add up to 1). }
\label{tab:GTEx}
\begin{center}
\begin{tabular}{l|c|c|c|c|c||c}
 & &  genotype & Sex & Platform & Unknown & Total \\
\toprule
 \multirow{3}{*}{Muscle}
       & Joint    & $17.09\%$  &  $2.03\%$  &  $2.56\%$ & $78.32\%$ & $16.44\%$ \\
       \cmidrule{2-7}
       & Individual    & $15.55\%$  &  $2.66\%$ &  $1.74\%$ & $80.05\%$ & $65.29\%$ \\
       \cmidrule{2-7}
       & Noise   &  &    &  & & $18.27\%$ \\
\midrule
 \multirow{3}{*}{Blood}
       & Joint    & $17.09\%$  &  $2.03\%$  &  $2.56\%$ & $78.32\%$ & $16.44\%$ \\
       \cmidrule{2-7}
       & Individual    & $14.05\%$  &  $0.65\%$ &  $0.90\%$ & $84.39\%$ & $63.56\%$ \\
       \cmidrule{2-7}
       & Noise   &  &    &  & & $20.00\%$ \\
\midrule
 \multirow{3}{*}{Skin}
       & Joint    & $17.09\%$  &  $2.03\%$  &  $2.56\%$ & $78.32\%$ & $16.44\%$ \\
       \cmidrule{2-7}
       & Individual    & $16.55\%$  &  $1.52\%$ &  $1.04\%$ & $80.89\%$ & $66.06\%$ \\
       \cmidrule{2-7}
       & Noise   &  &    &  & & $17.50\%$ \\
\bottomrule
\end{tabular}
\end{center}

\end{table}

\section{Discussion}\label{sec:discuss}

In this paper, we develop a supervised integrated factor analysis framework for reduction and integration of multi-view data.
It decomposes multiple related data sets into joint and individual structures, while incorporating covariate supervision through parametric or nonparametric models.
We investigate the identifiability of the model under two sets of conditions, the general conditions and the orthogonal conditions, each being useful in separate situations.
An efficient EM algorithm with some variants is devised to fit the model.
In particular, it is very easy to capture nonlinear relations between covariates and latent factors, and to incorporate variable selection of covariates.
The comprehensive simulation studies demonstrate the efficacy of the proposed methods.
With application to the GTEx data, we provide new insights into the genetic variation of a gene set across multiple tissues.

There are several directions for future research.
First of all, it is of potential interest to generalize the current framework to accommodate non-normal data.
Second, the model may be specially modified to capture partially joint structure pertaining to multiple but not all data sets.
This is especially relevant when multiple data sets are naturally grouped at the source level.
\textcolor{black}{Third, customized rank estimation methods need further investigation.}

\section*{Acknowledgements}
Gen Li was partially supported by the Calderone Junior Faculty Award by the Mailman School of Public Health at Columbia University.
Sungkyu Jung was partially supported by the National Science Foundation grant DMS-1307178.



\newpage
\appendix
\begin{center}
{\Large\bf Supplementary Materials for \\``Incorporating Covariates into Integrated Factor Analysis of Multi-View Data'' by Gen Li, and Sungkyu Jung}
\end{center}

\renewcommand\theequation{S.\arabic{equation}}
\setcounter{equation}{0}
\newcommand{\newcaption}{%
  \setlength{\abovecaptionskip}{-20pt}%
  \setlength{\belowcaptionskip}{0pt}%
  \caption}
\renewcommand{\thefigure}{S\arabic{figure}}
\renewcommand{\thetable}{S\arabic{table}}

\section{Identifiability}\label{suppsec:id}
In this section, we first give a few examples of unidentifiable models without any additional conditions other than the basic conditions (i.e., $\bV_k^T\bV_k=\bI$ and $\bSigma_k$ being diagonal with positive, distinct and decreasing eigenvalues for each $k=0,1,\cdots,K$).
Then we prove Proposition 2.1, the identifiability of the SIFA model under the general conditions.

\subsection{Examples of Unidentifiable Models}
By inspecting the log likelihood function (7) of the SIFA model, we can see that any two sets of model parameters are equivalent as long as they lead to the equal values of $\bmu_\star$ and $\bSigma_\star$, where
\bea\label{mu}
\bmu_\star&=&[\bdf_0(\bX),\bdf_1(\bX),\cdots,\bdf_K(\bX)](\bV_0,\bV_\star)^T;\\
\label{sigma}
\bSigma_\star&=&\bV_0\bSigma_0\bV_0^T+\bV_\star\bSigma_{\bF}\bV_\star^T+\bSigma_{\bE}.
\eea
In the following, we provide two examples of equivalent models.
One example involves models with different ranks, and the other example concerns models with the same set of ranks.

{\noindent\bf Example 1 (models with different ranks):}
Consider a very simple example with two primary data sets ($K=2$).
Suppose one model parameter set $\theta=\{\bdf_0(\cdot),\bdf_1(\cdot),\bdf_2(\cdot),\bV_{0},\bV_1,\bV_2,\bSigma_0,\bSigma_1,\bSigma_2,\\\sigma_1^2,\sigma_2^2\}$ satisfies the basic conditions  and has rank $r_0=1,r_1=r_2=2$.
Note that the combined loading matrix $(\bV_0,\bV_\star)$, where $\bV_\star=\blkdiag(\bV_1,\bV_2)$, may not have orthonormal columns.

One idea of generating an equivalent parameter set $\widehat{\theta}$ to $\theta$ is to intentionally mistake an individual component to a joint component and re-standardize everything accordingly. We define
\bes
\widetilde{\bV_0}=\left(\bV_0,\begin{pmatrix}
                           \bv_{1,1} \\
                           \0
                         \end{pmatrix}\right),
\widehat{\bV_1}=\bv_{1,2},
\widetilde{\bSigma_0}=\begin{pmatrix}
                           \bSigma_0 & \0 \\
                           \0 & \Sigma_{1,1}
                         \end{pmatrix},
\widehat{\bSigma_1}=\Sigma_{1,2},
\widetilde{\bdf_0}=(\bdf_0,f_{1,1}),
\widehat{\bdf_1}=f_{1,2},
\ees
where
$\bv_{1,1}$ and $\bv_{1,2}$ are the first and second columns of $\bV_1$, $\Sigma_{1,1}$ and $\Sigma_{1,2}$ are the first and second diagonal values of $\bSigma_1$, and $f_{1,1}$ and $f_{1,2}$ are the first and second components of $\bdf_{1}$.
In addition, let $\widehat{\bV_2}=\bV_2$, $\widehat{\bSigma_2}=\bSigma_2$, $\widehat{\bdf_2}=\bdf_2$, $\widehat{\sigma_1^2}=\sigma_1^2$, and $\widehat{\sigma_2^2}=\sigma_2^2$.

The matrix $\widetilde{\bV_0}$ is an intermediate loading matrix which may not have orthonormal columns (thus violating the basic conditions).
A new loading matrix $\widehat{\bV_0}$ and a new covariance matrix $\widehat{\bSigma_0}$ are obtained by the eigendecomposition of $\widetilde{\bV_0}\widetilde{\bSigma_0}\widetilde{\bV_0}^T$, 
where $\widehat{\bV_0}$ contains the eigenvectors and $\widehat{\bSigma_0}$ contains the eigenvalues.
As a result, $\widehat{\bV_0}$ and $\widehat{\bSigma_0}$ satisfy the basic conditions and \eqref{sigma} remains unchanged.
Since the column spaces of $\widetilde{\bV_0}$ and $\widehat{\bV_0}$ are  the same, we can easily obtain a $3\times3$ matrix $\bQ$ such that $\widetilde{\bV_0}=\widehat{\bV_0}\bQ$.
Correspondingly, we set $\widehat{\bdf_0}(\bX)=\widetilde{\bdf_0}(\bX)\bQ$.

As a result, we get a new parameter set $\widehat{\theta}=\{\widehat{\bdf_0}(\cdot),\widehat{\bdf_1}(\cdot),\widehat{\bdf_2}(\cdot),\widehat{\bV_{0}},\widehat{\bV_1},\widehat{\bV_2},\widehat{\bSigma_0},\widehat{\bSigma_1},\\\widehat{\bSigma_2},\widehat{\sigma_1^2},\widehat{\sigma_2^2}\}$.
It is easy to check that $\widehat{\bV_k}^T\widehat{\bV_k}=\bI$ and $\widehat{\bSigma_k}$'s are diagonal for $k=0,1,2$ (with a little care we can find an example where the resulting $\widehat{\bSigma_k}$ have distinct eigenvalues).
Namely, $\widehat{\theta}$ is a qualified parameter set with $r_0=2, r_1=1, r_2=2$.
According to the construction process, it is also trivial to see that $\theta$ and $\widehat{\theta}$ give exactly the same values in \eqref{mu} and \eqref{sigma}.
Therefore, $\theta$ and $\widehat{\theta}$ are equivalent but distinct model parameter sets.

{\noindent\bf Example 2 (models with the same set of ranks):}
Let us consider the same model parameter set $\theta$ as in Example 1.
Instead of padding the first column of $\bV_1$ with zeros and making it a joint loading vector, we pad the second column and make it a joint loading vector.
Specifically, let
\bes
\widetilde{\bV_0}'=\left(\bV_0,\begin{pmatrix}
                           \bv_{1,2} \\
                           \0
                         \end{pmatrix}\right),
\widehat{\bV_1}'=\bv_{1,1},
\widetilde{\bSigma_0}'=\begin{pmatrix}
                           \bSigma_0 & \0 \\
                           \0 & \Sigma_{1,2}
                         \end{pmatrix},
\widehat{\bSigma_1}'=\Sigma_{1,1},
\widetilde{\bdf_0}'=(\bdf_0,f_{1,2}),
\widehat{\bdf_1}'=f_{1,1},
\ees
and $\widehat{\bV_2}'=\bV_2$, $\widehat{\bSigma_2}'=\bSigma_2$, $\widehat{\bdf_2}'=\bdf_2$, $\widehat{\sigma_1^2}'=\sigma_1^2$, and $\widehat{\sigma_2^2}'=\sigma_2^2$.
Following exactly the same procedure as in Example 1, we can construct another equivalent parameter set $\widehat{\theta}'$ to $\theta$, with the ranks $r_0=2,r_1=1,r_2=2$.
Due to the transitive relation in the equivalent set, $\widehat{\theta}$ is equivalent to $\widehat{\theta}'$, and both parameter sets have the same  set of ranks.
We note that $\widehat{\theta}'$ is not identical to $\widehat{\theta}$ because $\widehat{\bV_1}=\bv_{1,2}\neq\bv_{1,1}=\widehat{\bV_1}'$.
As a result, we obtain two equivalent but distinct parameter sets with the same set of ranks.

In both examples, the parameter sets $\widehat{\theta}$ and $\widehat{\theta}'$ do not satisfy the general conditions. Next, we shall prove that under the general conditions, the parameter set is unique.

\subsection{Proof of Proposition 2.1}
\begin{proof}
Suppose $\theta=\{\bdf_0(\cdot),\bdf_k(\cdot),\bV_{0},\bV_k,\bSigma_0,\bSigma_k,\sigma_k^2;\ k=1,\cdots,K\}$ and  $\widehat{\theta}=\{\widehat{\bdf_0}(\cdot),\\\widehat{\bdf_k}(\cdot),\widehat{\bV_0},\widehat{\bV_k},\widehat{\bSigma_0},\widehat{\bSigma_k},\widehat{\sigma_k^2};\ k=1,\cdots,K\}$ are two equivalent model parameter sets that satisfy the basic conditions and the general conditions A1 and A2.
In the following, we will first prove $\widehat{\sigma_k^2}=\sigma_k^2$ for $k=1,\cdots, K$. Then we will prove that both parameter sets have the same set of   ranks, and the corresponding joint rank $r_0$ is minimal in the equivalent class.
Next, for each $k=0,1,\cdots,K$, we will prove $\widehat{\bV_k}=\bV_k$ (up to trivial column-wise sign changes) and $\widehat{\bSigma_k}=\bSigma_k$. Finally, we will prove $\widehat{\bdf_k}=\bdf_k$ (again, up to trivial sign changes).

{\bf Equal Variance:}
Let $\bSigma_\star$ and $\widehat{\bSigma_\star}$ denote the covariance matrices in \eqref{sigma} derived from the respective parameter sets.
Since $\bSigma_\star=\widehat{\bSigma_\star}$, all corresponding submatrices are equal.
In particular, we focus on the first $p_1\times p_1$ submatrix, and have $\bV_{0,1}\bSigma_0\bV_{0,1}^T+\bV_1\bSigma_1\bV_1^T+\sigma_1^2\bI=\widehat{\bV_{0,1}}\widehat{\bSigma_0}\widehat{\bV_{0,1}}^T+\widehat{\bV_1}\widehat{\bSigma_1}\widehat{\bV_1}^T+\widehat{\sigma_1^2}\bI$, or equivalently,
\be\label{eqvar}
(\bV_{0,1},\bV_1)\begin{pmatrix}
                   \bSigma_{0} & \0 \\
                   \0 & \bSigma_{1}
                 \end{pmatrix}
                \begin{pmatrix}
                   \bV_{0,1}^T \\
                   \bV_{1}^T
                 \end{pmatrix} +\sigma_1^2\bI=
(\widehat{\bV_{0,1}},\widehat{\bV_1})\begin{pmatrix}
                   \widehat{\bSigma_{0}} & \0 \\
                   \0 & \widehat{\bSigma_{1}}
                 \end{pmatrix}
                \begin{pmatrix}
                   \widehat{\bV_{0,1}}^T \\
                   \widehat{\bV_{1}}^T
                 \end{pmatrix} +\widehat{\sigma_1^2}\bI.
\ee
From Condition A2, we know $r_0+r_1<p_1$ and $\widehat{r_0}+\widehat{r_1}<p_1$, where $r_k$ and $\widehat{r_k}$ are the numbers of columns in $\bV_k$ and $\widehat{\bV_k}$ respectively ($k=0,1\cdots,K$).
Thus, the first term on both sides of the equation is a low-rank non-negative definite matrix.
It is easy to see the smallest eigenvalue of the left-hand side (LHS) matrix is $\sigma_1^2$ and the smallest eigenvalue of the right-hand side (RHS) matrix is $\widehat{\sigma_1^2}$.
Therefore, we have $\sigma_1^2=\widehat{\sigma_1^2}$.
Similar argument for the $k$th diagonal submatrix leads to $\sigma_k^2=\widehat{\sigma_k^2}, \ k=1,\cdots,K$.

{\bf Equal Ranks:}
Taking out the second term from both sides of \eqref{eqvar}, we have
\bes
(\bV_{0,1},\bV_1)\begin{pmatrix}
                   \bSigma_{0} & \0 \\
                   \0 & \bSigma_{1}
                 \end{pmatrix}
                \begin{pmatrix}
                   \bV_{0,1}^T \\
                   \bV_{1}^T
                 \end{pmatrix}=
(\widehat{\bV_{0,1}},\widehat{\bV_1})\begin{pmatrix}
                   \widehat{\bSigma_{0}} & \0 \\
                   \0 & \widehat{\bSigma_{1}}
                 \end{pmatrix}
                \begin{pmatrix}
                   \widehat{\bV_{0,1}}^T \\
                   \widehat{\bV_{1}}^T
                 \end{pmatrix}.
\ees
Since $(\bV_{0,1},\bV_1)$ and $(\widehat{\bV_{0,1}},\widehat{\bV_1})$ both have full column ranks according to Condition A1 and A2, and the diagonal values of $\blkdiag(\bSigma_0,\bSigma_1)$ and $\blkdiag(\widehat{\bSigma_0},\widehat{\bSigma_1})$ are positive, the LHS matrix and the RHS matrix have ranks $r_0+r_1$ and $\widehat{r_0}+\widehat{r_1}$, respectively.
Consequently, we get $k$ equations: $r_0+r_k=\widehat{r_0}+\widehat{r_k}$ for $k=1,\cdots,K$.
In addition, from $\bSigma_\star=\widehat{\bSigma_\star}$, we get another equation: $r_0+\sum_{k=1}^K r_k=\widehat{r_0}+\sum_{k=1}^K \widehat{r_k}$.
Solving these $K+1$ equations, we get $r_k=\widehat{r_k}$, for $k=0,1,\cdots,K$.

To see that the $r_0$ of $\theta$ is actually minimal in the equivalent class (in which parameter sets do not necessarily satisfy the general conditions), we can look at an off-diagonal block matrix of $\bSigma_\star$.
For example, the submatrix consisting of the $(p_1+1)$th to the $(p_1+p_2)$th columns and the first to the $p_1$th rows is $\bV_{0,1}\bSigma_0\bV_{0,2}^T$. It has rank $r_0$.
For any parameter set in the equivalent class, the derived submatrix should be identical with $\bV_{0,1}\bSigma_0\bV_{0,2}^T$ and thus also have rank $r_0$.
Consequently, it must have no fewer than $r_0$ joint loadings.
Namely, $r_0$ is minimal in the equivalent class.

{\bf Equal Loading and Covariance:}
In order to prove the loading matrices are equal, we first look at the off-diagonal block matrices of $\bSigma_\star$ and $\widehat{\bSigma_\star}$.
In particular, we have
\be\label{temp}
\bV_{0,k_1}\bSigma_0\bV_{0,k_2}^T=\widehat{\bV_{0,k_1}}\widehat{\bSigma_0}\widehat{\bV_{0,k_2}}^T, \ k_1\neq k_2\in\{1,\cdots,K\}.
\ee
Based on Condition A1, the LHS and RHS matrices both have rank $r_0$, and the column spaces are $\col(\bV_{0,k_1})$ and $\col(\widehat{\bV_{0,k_1}})$ respectively. Thus we have $\col(\bV_{0,k})=\col(\widehat{\bV_{0,k}})$ for $k=1,\cdots,K$.

Then we look at the diagonal block matrices.
In particular, we have
\be\label{eqlast}
\bV_{0,1}\bSigma_0\bV_{0,1}^T+\bV_1\bSigma_1\bV_1^T=\widehat{\bV_{0,1}}\widehat{\bSigma_0}\widehat{\bV_{0,1}}^T+\widehat{\bV_1}\widehat{\bSigma_1}\widehat{\bV_1}^T.
\ee
Let $\bC=\bV_{0,1}\bSigma_0\bV_{0,1}^T-\widehat{\bV_{0,1}}\widehat{\bSigma_0}\widehat{\bV_{0,1}}^T$.
From the previous discussion, we know $\col(\bC)\subseteq\col(\bV_{0,1})$.
Plugging $\bC$ into \eqref{eqlast}, we have
\bes
\widehat{\bV_1}\widehat{\bSigma_1}\widehat{\bV_1}^T=\bV_1\bSigma_1\bV_1^T+\bC.
\ees
According to Corollary 18.5.5 in \cite{harvillematrix}, the rank of the RHS matrix is bounded below by $\rank(\bV_1\bSigma_1\bV_1^T)+\rank(\bC)-2c$, where $c$ is the dimension of $\col(\bV_1\bSigma_1\bV_1^T)\cap\col(\bC)$.
Since $\col(\bV_1\bSigma_1\bV_1^T)=\col(\bV_1)$, $\col(\bC)\subseteq\col(\bV_{0,1})$, and $\col(\bV_1)\cap\col(\bV_{0,1})=\{\0\}$ (according to Condition A2), we know $c=0$.
Therefore, we have
\bes
\rank(\widehat{\bV_1}\widehat{\bSigma_1}\widehat{\bV_1}^T)=\rank(\bV_1\bSigma_1\bV_1^T+\bC)\geq\rank(\bV_1\bSigma_1\bV_1^T)+\rank(\bC)=r_1+\rank(\bC).
\ees
In combination with the fact that $\rank(\widehat{\bV_1}\widehat{\bSigma_1}\widehat{\bV_1}^T)=\widehat{r_1}=r_1$, we get $\rank(\bC)=0$, i.e., $\bC=\0$.

Immediately, we have  $\widehat{\bV_k}\widehat{\bSigma_k}\widehat{\bV_k}^T=\bV_k\bSigma_k\bV_k^T$ and
$\widehat{\bV_{0,k}}\widehat{\bSigma_0}\widehat{\bV_{0,k}}^T=\bV_{0,k}\bSigma_0\bV_{0,k}^T$ for $k=1,\cdots,K$. Combining with \eqref{temp}, we also have $\widehat{\bV_{0}}\widehat{\bSigma_0}\widehat{\bV_{0}}^T=\bV_{0}\bSigma_0\bV_{0}^T$.
Under the basic conditions, $\bV_k^T\bSigma_k\bV_k$ and $\widehat{\bV_k}^T\widehat{\bSigma_k}\widehat{\bV_k}$  ($k=0,1,\cdots,K$) are both in the form of eigendecomposition. From the uniqueness of eigendecomposition, we conclude that
\bes
\bV_k=\widehat{\bV_k}\bS_k,\ \bSigma_k=\widehat{\bSigma_k},\ k=0,1,\cdots,K,
\ees
where $\bS_k$ is a diagonal matrix (with dimension compatible with $\widehat{\bV_k}$) whose diagonal values are either $1$ or $-1$.
For an eigenvector, the sign change is trivial.
As discussed in the main article, one could easily fix the sign of each eigenvector by setting the first nonzero entry to be positive.
Without loss of generality, we assume $\bS_k$ is the identity matrix.

{\bf Equal Functions:}
We plug in $\bV_0=\widehat{\bV_0}$ and $\bV_\star=\widehat{\bV_\star}$ in \eqref{mu}.
As a result, we obtain that $\bdf_k(\bX)-\widehat{\bdf_k}(\bX)=\0$ holds for any $\bX$ $(k=0,1,\cdots,K)$. Namely, $\bdf_k=\widehat{\bdf_k}$, $k=0,1,\cdots,K$.
\end{proof}

\section{Algorithm Details}\label{suppsec:alg}
In this section, we provide a detailed description of the EM algorithm for fitting an SIFA model.
The M steps for different sets of identifiability conditions are slightly different, while the E step is universal.
In the following, we will derive the conditional distribution of the latent variables, and then derive the specifics of the M steps under different conditions.
We will use the notations used in the main article wherever applicable.

\subsection{E Step}
Assume $\theta^{(l)}$ is the parameter set estimated from the $l$th iteration.
We focus on the estimation in the $(l+1)$th iteration.
For simplicity, we shall drop the superscript when it does not cause any confusion.
Under the normal assumption, the conditional distribution of $\by_{\star(i)}$ (a column vector of the $i$th row of $\bY_\star$) given $\bu_{0(i)}$ and $\bu_{\star(i)}$ (column vectors of the $i$th row of $\bU_0$ and $\bU_\star$) is
\bes
\by_{\star(i)}|\bu_{0(i)},\bu_{\star(i)} \ \sim\  \norm (\bmu_{(i)},\bSigma_{\bE}),
\ees
where $\bmu_{(i)}$ is a column vector of the $i$th row of $\bU_0\bV_0^T+\bU_\star\bV_\star^T$.
The marginal distributions of $\bu_{0(i)}$ and $\bu_{\star(i)}$ are
\bes
\bu_{0(i)}\ &\sim& \ \norm\left({\bdf_0(\bx_{(i)})}^T,\bSigma_0\right),\\
\bu_{\star(i)} \ &\sim& \ \norm\left([\bdf_1(\bx_{(i)}),\cdots,\bdf_K(\bx_{(i)})]^T,\bSigma_{\bF}\right),
\ees
where $\bdf_k(\bx_{(i)})$ is the $i$th row of $\bdf_k(\bX)$ for $k=0,1,\cdots,K$.
With some basic algebraic calculation, we can easily get the marginal distribution of $\by_{\star(i)}$ and the joint distribution of $\by_{\star(i)},\bu_{0(i)},\bu_{\star(i)}$ as a normal distribution. To avoid overcomplicated notations, we will not present the intermediate results here.

Subsequently, we obtain the conditional distribution of $\bu_{0(i)},\bu_{\star(i)}$ given $\by_{\star(i)}$, which is a normal distribution.
In particular, the matrix form of the conditional mean is
\be\label{mean}
\Ex(\bU_0,\bU_\star|\bY_\star) =\left[\bdf_0(\bX),\bdf_1(\bX),\cdots,\bdf_K(\bX)\right]+(\bY_\star-\bmu_\star)\bSigma_\star^{-1}(\bV_0\bSigma_0,\bV_\star\bSigma_{\bF})
\ee
where $\bSigma_\star$ is the marginal covariance matrix of $\by_{\star(i)}$, which has the form \eqref{sigma}.
The conditional covariance matrix of each row of $(\bU_0,\bU_\star)$ given $\bY_\star$ is
\be\label{var}
\begin{pmatrix}
  \bSigma_0 & \0 \\
  \0 & \bSigma_{\bF}
\end{pmatrix}-
\begin{pmatrix}
  \bSigma_0 & \0 \\
  \0 & \bSigma_{\bF}
\end{pmatrix}
\begin{pmatrix}
  \bV_0^T \\
  \bV_\star^T
\end{pmatrix}\bSigma_\star^{-1}
(\bV_0,\bV_\star)
\begin{pmatrix}
  \bSigma_0 & \0 \\
  \0 & \bSigma_{\bF}
\end{pmatrix}.
\ee
Using the Woodbury matrix identity, we can further simplify $(\bV_0,\bV_\star)^T\bSigma_\star^{-1}(\bV_0,\bV_\star)$ as
\[
\begin{pmatrix}
  \bV_0^T \\
  \bV_\star^T
\end{pmatrix}\bSigma_\star^{-1}
(\bV_0,\bV_\star)
=
\bDelta-\bDelta\left[\begin{pmatrix}
  \bSigma_0^{-1} & \0 \\
  \0 & \bSigma_{\bF}^{-1}
\end{pmatrix}
+\bDelta\right]^{-1}\bDelta,
\]
where $\bDelta=(\bV_0,\bV_\star)^T\bSigma_{\bE}^{-1}(\bV_0,\bV_\star)$.
Namely, at most we need to invert a $\sum_{k=0}^K r_k \times \sum_{k=0}^K r_k$ matrix, which is computationally feasible even for high dimensional data.
In particular, under the orthogonal conditions, $\bDelta$ is a diagonal matrix: $$\bDelta=\blkdiag\left(\sum_{k=1}^K(\sigma_k^{-2}/K)\bI_{r_0},\sigma_1^{-2}\bI_{r_1},\cdots,\sigma_K^{-2}\bI_{r_K}\right).$$
The computation is further simplified.
All in all, the conditional distribution of the latent variables is fully derived.
It is straightforward to calculate any conditional moment functions of $(\bU_0,\bU_\star)$.
Therefore, we are ready for the M step.

\subsection{M Step (Under the General Conditions)}
In the M step, we maximize the conditional expectation of the joint log likelihood of the observed data and the latent variables. It can be separated into two sets of optimization problems: maximizing the conditional expectation of the marginal likelihood of $\bU_k$'s ($k=0,\cdots,K$), and maximizing the conditional expectation of the conditional likelihood of $\bY_\star$ given $(\bU_0,\bU_\star)$.
\vskip.2in
{\bf First Optimization:}
The first set of optimization is relatively straightforward.
Essentially, we need to solve the following problem
\bes
\min\limits_{\bdf_k(\cdot),\bSigma_k} \  n\log|\bSigma_k|+\mathbb{E}_{\bU_k|\bY_\star}\left\{\tr\left[(\bU_k-\bdf_k(\bX))\bSigma_k^{-1}(\bU_k-\bdf_k(\bX))^T\right]\right\},
\ees
for $k=0,1,\cdots, K$.
Since we assume $\bSigma_k$ is diagonal and $\bdf_k=(f_{k,1},\cdots,f_{k,r_k})$ contains $r_k$ separate functions, the above problem can be further separated as
\bea\label{origU}
\min\limits_{f_{k,j}(\cdot)} \  \mathbb{E}_{\bU_k|\bY_\star}\left\{\tr\left[(\bu_{k,j}-f_{k,j}(\bX))(\bu_{k,j}-f_{k,j}(\bX))^T\right]\right\},\quad j=1,\cdots,r_k;
\eea
and
\bea\label{eq:2}
\min\limits_{\bSigma_k} \  n\log|\bSigma_k|+\mathbb{E}_{\bU_k|\bY_\star}\left\{\tr\left[(\bU_k-\bdf_k(\bX))\bSigma_k^{-1}(\bU_k-\bdf_k(\bX))^T\right]\right\}.
\eea

\textcolor{black}{After adding and subtracting some constant terms, \eqref{origU} is equivalent to the least square problem:
\bea\label{eq:1}
\min_{f_{k,j}(\cdot)}\ \|\mathbb{E}(\bu_{k,j}|\bY_\star)-f_{k,j}(\bX)\|_\mathbb{F}^2,
\eea
for $j=1,\cdots,r_k$.
The optimization problem \eqref{eq:1} has been well studied in the literature, and can be readily solved via off-the-shelf solvers.
More specifically, if $f_{k,j}$ is a linear function, it becomes ordinary least squares
\bea\label{ols}
\min_{\bbeta_{k,j}} \|\mathbb{E}(\bu_{k,j}|\bY_\star)-\bX\bbeta_{k,j}\|_\mathbb{F}^2,
\eea
which has an explicit solution $\widehat{\bbeta_{k,j}}=(\bX^T\bX)^{-1}\bX\mathbb{E}(\bu_{k,j}|\bY_\star)$.
If $f_{k,j}$ is nonparametric, \eqref{eq:1} can be solved using kernel methods or spline-based methods \citep[][]{hastie1990generalized,fan1996local,hollander2013nonparametric}.}

\textcolor{black}{
It is also straightforward to incorporate variable selection in \eqref{eq:1} if $f_{k,j}$ is additive.
For example, if $f_{k,j}$ is linear as in \eqref{ols}, we can add a sparsity-inducing penalty, such as the LASSO penalty \citep{tibshirani1996regression,efron2004least}, to the objective function \eqref{ols}, and solve the penalized least squares problem to get a sparse estimate of $\bbeta_{k,j}$.
Tuning parameters can be selected adaptively.
We particularly apply this procedure to the Berkeley Growth Study example in Section \ref{suppsec:growth}.
The results are more interpretable after variable selection.
More generally, if $f_{k,j}$ is some nonparametric additive function, one may replace \eqref{eq:1} with the methods for estimating sparse additive models proposed in the literature \citep[cf.][]{lafferty2008rodeo,ravikumar2009sparse}.
All in all, incorporating variable selection into \eqref{eq:1} is a well studied problem. With variable selection, we can accommodate high dimensional covariates and identify important covariates for each latent factor.}

Once \eqref{eq:1} is solved, the optimization problem \eqref{eq:2} can be solved explicitly as
\be\label{eq5}
\widehat{\bSigma_k}={1\over n}\diag\left\{\Ex_{{\bU_k|\bY_\star}}\left[\left(\bU_k-\widehat{\bdf_k}(\bX)\right)^T\left(\bU_k-\widehat{\bdf_k}(\bX)\right)\right]\right\}, \ k=0,1,\cdots,K
\ee
where $\widehat{\bdf_k}=(\widehat{f_{k,1}},\cdots,\widehat{f_{k,r_k}})$ is the optimizer of \eqref{eq:1}.

\vskip.2in
{\bf Second Optimization:}
The second set of optimization can be written as
\be\label{se}
\min\limits_{\bV_0,\bV_\star,\sigma_1^2,\cdots,\sigma_K^2} \ \sum_{k=1}^K \left[np_k\log\sigma_k^2 +  \sigma_k^{-2}\Ex_{\bU_0,\bU_k|\bY_\star}\|\bY_k-\bU_k\bV_{k}^T-\bU_0\bV_{0,k}^T\|_\mathbb{F}^2\right],
\ee
where $\bV_0$ and $\bV_\star$ satisfy the basic conditions and the general conditions.
Under those conditions, there are no closed-form expressions for $\bV_0$ and $\bV_k$ ($k=1,\cdots,K$).
As a remedy, we solve the above problem with respect to $\bV_k$ and $\bV_0$ sequentially.
Fixing $\bV_0$, the optimization of $\bV_k$ becomes
\bea\nonumber
&&\min\limits_{\bV_k:\bV_k^T\bV_k=\bI} \ \Ex_{\bU_0,\bU_k|\bY_\star} \|\bY_k-\bU_0\bV_{0,k}^T-\bU_k\bV_k^T\|_\mathbb{F}^2\\
\nonumber
&\Leftrightarrow&
\min\limits_{\bV_k:\bV_k^T\bV_k=\bI} \ \Ex_{\bU_0,\bU_k|\bY_\star} \left\{-2\tr\left[\left(\bY_k-\bU_0\bV_{0,k}^T\right)^T\bU_k\bV_k^T\right]\right\}\\
\nonumber
&\Leftrightarrow&
\max\limits_{\bV_k:\bV_k^T\bV_k=\bI} \ \tr \left[\Ex\left(\bY_k^T\bU_k-\bV_{0,k}\bU_0^T\bU_k|\bY_\star\right)\bV_k^T\right]\\
\label{eq:3}
&\Leftrightarrow&
\max\limits_{\bV_k:\bV_k^T\bV_k=\bI} \ \tr\left\{ \left[\bY_k^T\Ex(\bU_k|\bY_\star)-\bV_{0,k}\Ex(\bU_0^T\bU_k|\bY_\star)\right]\bV_k^T\right\}.
\eea
By the singular value decomposition (SVD), we have $\bY_k^T\Ex(\bU_k|\bY_\star)-\bV_{0,k}\Ex(\bU_0^T\bU_k|\bY_\star)=\bL\bD\bR^T$, where $\bL$ ($\bR$) contains $r_k$ left (right) singular vectors, and $\bD$ is a diagonal matrix with $r_k$ positive singular values on the diagonal.
The object function in \eqref{eq:3} satisfies the following relations
\bes
\tr(\bL\bD\bR^T\bV_k^T)=\tr(\bD\bR^T\bV_k^T\bL)\leq\tr(\bD),
\ees
where ``$=$" holds for the second inequality if and only if $\bV_k=\bL\bR^T$.
This is because $\bV_k\bR$ and $\bL$ both have $r_k$ orthonormal columns.
The diagonal values of their inner product are no larger than 1.
Since $\bD$ has $r_k$ positive diagonal values, $\tr(\bD\bR^T\bV_k^T\bL)$ is maximized if and only if $\bV_k\bR=\bL$.
Namely, the optimal solution for \eqref{eq:3} is
\be\label{eq:9}
\widehat{\bV_k}=\bL\bR^T.
\ee

When $\bV_k$'s are held fixed, the optimization of $\bV_0$ becomes
\be\label{eq:4}
\min\limits_{\bV_0: \bV_0^T\bV_0=\bI} \ \sum_{k=1}^K   \sigma_k^{-2}\Ex_{\bU_0,\bU_k|\bY_\star}\|\bY_k-\bU_k\bV_{k}^T-\bU_0\bV_{0,k}^T\|_\mathbb{F}^2.
\ee
There is no closed-form solution for this constrained optimization problem.
As a remedy, we use a relax-and-retrieve strategy to approximately solve the problem: first relax the orthogonality constraint and derive a closed-form solution for $\bV_0$, and then retrieve the orthogonality through the eigendecomposition of $\bV_0\bSigma_0\bV_0^T$.
Without the constraint, the solution of \eqref{eq:4} is
\be\label{eq:5}
\widetilde{\bV_{0,k}}= \left[\bY_k^T\mathbb{E}(\bU_0|\bY_\star)-\bV_k\mathbb{E}(\bU_k^T\bU_0|\bY_\star)\right]\left[\mathbb{E}(\bU_0^T\bU_0|\bY_\star)\right]^{-1},\ k=1,\cdots,K.
\ee
Subsequently, we orthogonalize the columns in $\widetilde{\bV_0}$ through the eigendecomposition of $\widetilde{\bV_0}\widehat{\bSigma_0}\widetilde{\bV_0}^T$.
In particular, we set the columns of $\widehat{\bV_0}$ to be the eigenvectors, and replace the original diagonal values of $\widehat{\bSigma_0}$ with the eigenvalues of $\widetilde{\bV_0}\widehat{\bSigma_0}\widetilde{\bV_0}^T$.
This step will not change the value of $\bSigma_\star$ in \eqref{sigma}.
A similar approach has been adopted in \cite{li2016superviseda}.

Once $\widehat{\bV_0}$ and $\widehat{\bV_k}$ are obtained, it is straightforward to optimize \eqref{se} to obtain the estimator of $\sigma_k^2$ as
\be\label{eq:7}
\widehat{\sigma_k^2}=\mathbb{E}_{\bU_0,\bU_k|\bY_\star}\|\bY_k-\bU_k\widehat{\bV_{k}}^T-\bU_0\widehat{\bV_{0,k}}^T\|_\mathbb{F}^2,\quad k=1,\cdots,K.
\ee

\subsection{M Step (Under the Orthogonal Conditions)}
Under the orthogonal conditions, the first set of optimization remains the same with those under the general conditions.
We only show a variant of the algorithm for the second set of optimization.

With the orthogonal constraint $(\sqrt{K}\bV_{0,k},\bV_k)^T(\sqrt{K}\bV_{0,k},\bV_k)=\bI$ for $k=1,\cdots,K$, the original optimization problem \eqref{se} can be separated into $K$ parts and solved in parallel.
In particular, $\bV_{0,k}$ and $\bV_k$ can be estimated together via solving  the following optimization under the orthogonal constraint
\bea\nonumber
&&\min\limits_{\bV_{0,k},\bV_k} \  \Ex_{\bU_0,\bU_k|\bY_\star}\|\bY_k-\bU_k\bV_{k}^T-\bU_0\bV_{0,k}^T\|_\mathbb{F}^2\\
\nonumber
&\Leftrightarrow&
\min\limits_{\bV_{0,k},\bV_k} \ \Ex_{\bU_0,\bU_k|\bY_\star}\|\bY_k-({1\over \sqrt{K}}\bU_0,\bU_k)(\sqrt{K}\bV_{0,k},\bV_{k})^T\|_\mathbb{F}^2\\
\label{eq:6}
&\Leftrightarrow&
\min\limits_{\bV_{0,k},\bV_k} \ \|\bY_k-\left({1\over \sqrt{K}}\Ex(\bU_0|\bY_\star),\Ex(\bU_k|\bY_\star)\right)(\sqrt{K}\bV_{0,k},\bV_{k})^T\|_\mathbb{F}^2.
\eea
The above problem is exactly in the form of an orthogonal Procrustes problem \citep{gower2004procrustes}.
Thus the unique optimal solution for \eqref{eq:6} is
\be\label{eq:8}
(\sqrt{K}\widehat{\bV_{0,k}},\widehat{\bV_{k}})=\bL\bR^T,
\ee
where $\bL$ and $\bR$ contain the $r_0+r_k$ left and right singular vectors of $\bY_k^T\left({1/\sqrt{K}}\Ex(\bU_0|\bY_\star),\Ex(\bU_k|\bY_\star)\right)$, respectively.
Subsequently, we obtain $\widehat{\bV_0}$ by concatenating the respective $\widehat{\bV_{0,k}}$'s.
Once the loading matrices are estimated, we use the same estimator as before to estimate $\sigma_k^2$.

\subsection{Flow Chart}
\textcolor{black}{We summarize the EM algorithms for model fitting under different identifiability conditions in Algorithm \ref{alg1} and Algorithm \ref{alg2}.
We emphasize that under the orthogonal conditions, the computation is extremely efficient.
There is no need for sequential optimization or approximation.}

\begin{algorithm}[h]
\textcolor{black}{
\caption{The EM algorithm under the {\em general conditions}}\label{alg1}
\begin{algorithmic}
\State Initialize $\theta_0$, possibly from JIVE or PCA estimates ;
\While {Estimation has not reached convergence}\\
{\quad \bf E Step:}
\State Calculate the conditional mean \eqref{mean} and conditional variance \eqref{var}\\
{\quad\bf M Step:}
\State Estimate $f_{k,j}$ ($j=1,\cdots,r_k;k=0,1,\cdots,K$) by solving \eqref{eq:1};
\State Estimate $\bSigma_k$ ($k=0,1,\cdots,K$) via \eqref{eq5};
\State Cycle through the following steps (for one round or until convergence):
\bi
\State Estimate $\bV_k$ ($k=1,\cdots,K$) while fixing $\bV_0$ via \eqref{eq:9};
\State Obtain an intermediate estimate of $\bV_{0,k}$ while fixing $\bV_k$ ($k=1,\cdots,K$) via \eqref{eq:5};
\State Normalize the intermediate estimate of $\bV_0$ and $\bSigma_0$ together through SVD;
\ei
\State Estimate $\sigma^2_k$ ($k=1,\cdots,K$) via \eqref{eq:7};
\EndWhile
\end{algorithmic}}
\end{algorithm}

\begin{algorithm}[h]
\textcolor{black}{
\caption{The EM algorithm under the {\em orthogonal conditions}}\label{alg2}
\begin{algorithmic}
\State Initialize $\theta_0$, possibly from JIVE or PCA estimates ;
\While {Estimation has not reached convergence}\\
{\quad \bf E Step:}
\State Calculate the conditional mean \eqref{mean} and conditional variance \eqref{var}\\
{\quad\bf M Step:}
\State Estimate $f_{k,j}$ ($j=1,\cdots,r_k;k=0,1,\cdots,K$) by solving \eqref{eq:1};
\State Estimate $\bSigma_k$ ($k=0,1,\cdots,K$) via \eqref{eq5};
\State Estimate $\bV_{0,k}$ and $\bV_k$ ($k=1,\cdots,K$) via \eqref{eq:8}
\State Estimate $\sigma^2_k$ ($k=1,\cdots,K$) via \eqref{eq:7};
\EndWhile
\end{algorithmic}}
\end{algorithm}

\section{Likelihood Cross Validation for Refining Rank Estimation}\label{suppsec:rank}

%
%

\textcolor{black}{
In the SIFA model, a set of the joint rank $r_0$ and the individual ranks $r_k$ ($k=1,\cdots,K$) is treated as a tuning parameter, denoted by $\br=(r_0,r_1,\cdots,r_K)$.
Given a collection of candidate rank sets, we introduce  an $N$-fold LCV method to select the best rank set.
More specifically, we first randomly split the samples across all data sets into $N$ groups.
In each cross validation run, one block of samples is treated as testing samples, and the other samples are used as training samples.
Suppose we have $m$ candidate rank sets.
In each run, we fit $m$ models under different ranks using the training samples.
Then we calculate the log likelihood value of the testing samples for each fitted model, using Eqn (7) in the main paper.
The negative log likelihood values are used as LCV scores, where a smaller value is more desired.
We repeat the procedure $N$ times and average the LCV scores for each candidate rank set.
The rank set corresponds to the smallest score is selected.
We remark that an $N$-fold LCV study on $M$ candidate rank sets requires fitting the SIFA model $M*N$ times.
This may be computationally intensive when $N$ or $M$ is large.
Therefore, in practice, when the data dimension is large, we recommend estimating the ranks with the two-step procedure first, and then using the LCV approach to refine the result by searching among the neighbors of the crude estimate.
}

{\color{black}
To demonstrate the efficacy of the LCV approach, we consider a couple of simulated examples.
In particular, data are generated from Setting 1 and 2, respectively, in Section 3.1 of the main paper, with the true ranks to be $r_0=2, r_1=r_2=3$.
We consider 9 rank sets: $(r_0,r_1,r_2)\in\{(1,2,2),(2,2,2),(3,2,2),(1,3,3),(2,3,3),(3,3,3),\\(3,4,3),(3,4,4),(4,4,4)\}$, in the neighborhood of the true rank set.
The 10-fold LCV scores (i.e., negative log likelihood values) are shown in Figure \ref{fig:set2} (for Setting 2) and Figure \ref{fig:set1} (for Setting 1), respectively.

When the data are generated from Setting 2 (i.e., the SIFA-A model), the LCV approach correctly selects the true rank set.
We remark that the sets with large ranks all have relatively small LCV scores, so there may be the risk of overestimation.
When the data are generated from Setting 1 (i.e., the JIVE model), the LCV approach does not correctly select the ranks. The CV scores for different rank sets are quite similar.
This may be due to the lack of relevant information from auxiliary covariates.
A more powerful and customized rank estimation procedure requires further investigation.
\begin{figure}[t]
\begin{center}
\includegraphics[width=4in]{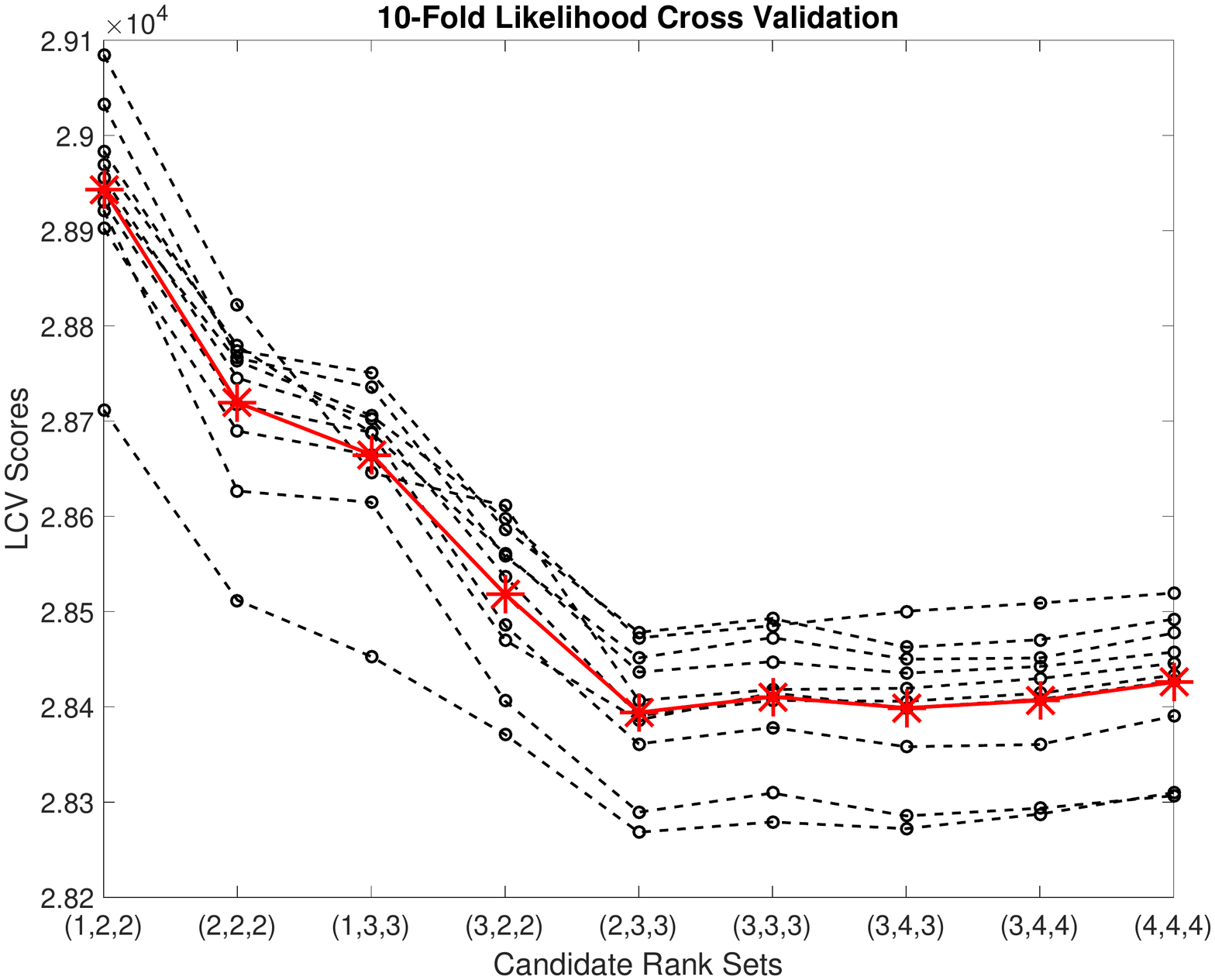}
\end{center}
\newcaption{Setting 2: The LCV scores for 10-fold cross validation on 9 candidate rank sets. Each dashed line with circles contains corresponds to the LCV scores (negative log likelihood values) in one fold. The solid line with stars contains the average LCV scores for different rank sets. }
\label{fig:set2}
\end{figure}
\begin{figure}[t]
\begin{center}
\includegraphics[width=4in]{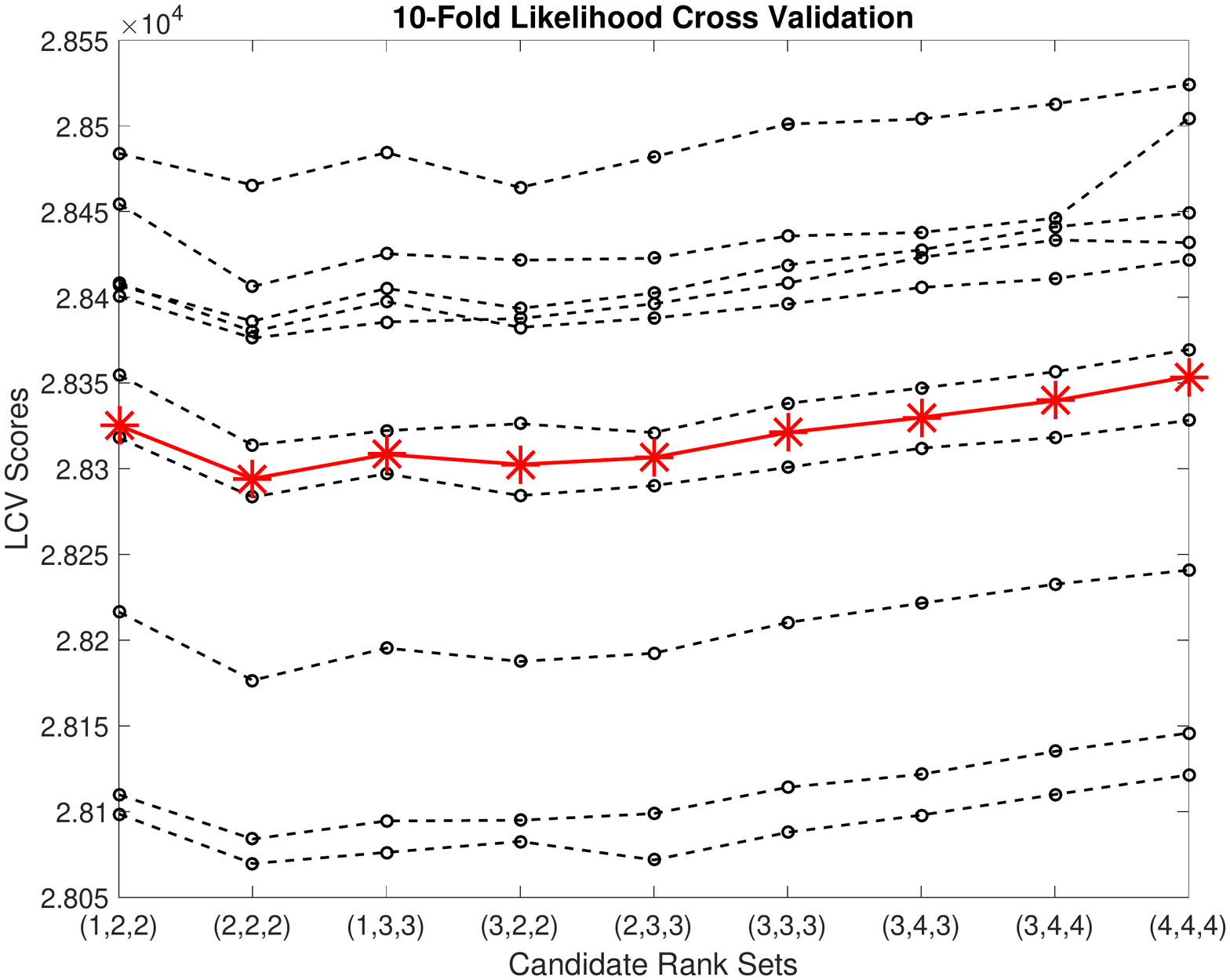}
\end{center}
\newcaption{Setting 1: The LCV scores for 10-fold cross validation on 9 candidate rank sets. Each dashed line with circles contains corresponds to the LCV scores (negative log likelihood values) in one fold. The solid line with stars contains the average LCV scores for different rank sets. }
\label{fig:set1}
\end{figure}

}

\section{Additional Simulation Studies}\label{suppsec:sim}
\subsection{Nonlinear Covariate Functions}
\textcolor{black}{
To investigate nonparametric model fitting under nonlinear covariate functions, we consider the following simulation setting.
In particular, the parameters and the ranks are set in the same way as in Setting 2 in the main manuscript, except  there is only one covariate ($q=1$) and it is related to different latent factors in highly nonlinear fashions.
\bi
\item {\bf Setting 4} (SIFA-A Model with nonlinear relations): The factors are generated from $\bU_k=\bdf_k(X)+\bF_k$ for $k=0,1,2$, where $X$ represents the univariate covariate. The univariate functions in $\bdf_0(\cdot),\bdf_1(\cdot),\bdf_2(\cdot)$ contain sine, cosine, quadratic, and cubic functions. Other parameters are generated in compliance with the general conditions.
\ei
}

\textcolor{black}{
We focus on SIFA-A models with nonparametric estimation and linear estimation. The nonparametric estimation is achieved using a kernel regression method.
In addition, we also consider the JIVE model without incorporating the covariate.
The comparison results are shown in Figure \ref{fig:nonlinear}.
In terms of the estimation accuracy for separate loadings and the combined loading subspace, it is apparent that using the kernel regression method in SIFA-A achieves the best result.
The SIFA-A model with misspecified linear covariate functions has comparable performance with JIVE.
For the low-rank structure recovery accuracy, both SIFA-A models significantly outperform the JIVE model, with the nonparametric version being slightly better than the linear version.
Overall, incorporating the auxiliary covariate into the model improves parameter estimation and dimension reduction. SIFA is also robust against covariate function misspecification.
\begin{figure}[!]
\begin{center}
\includegraphics[width=6in]{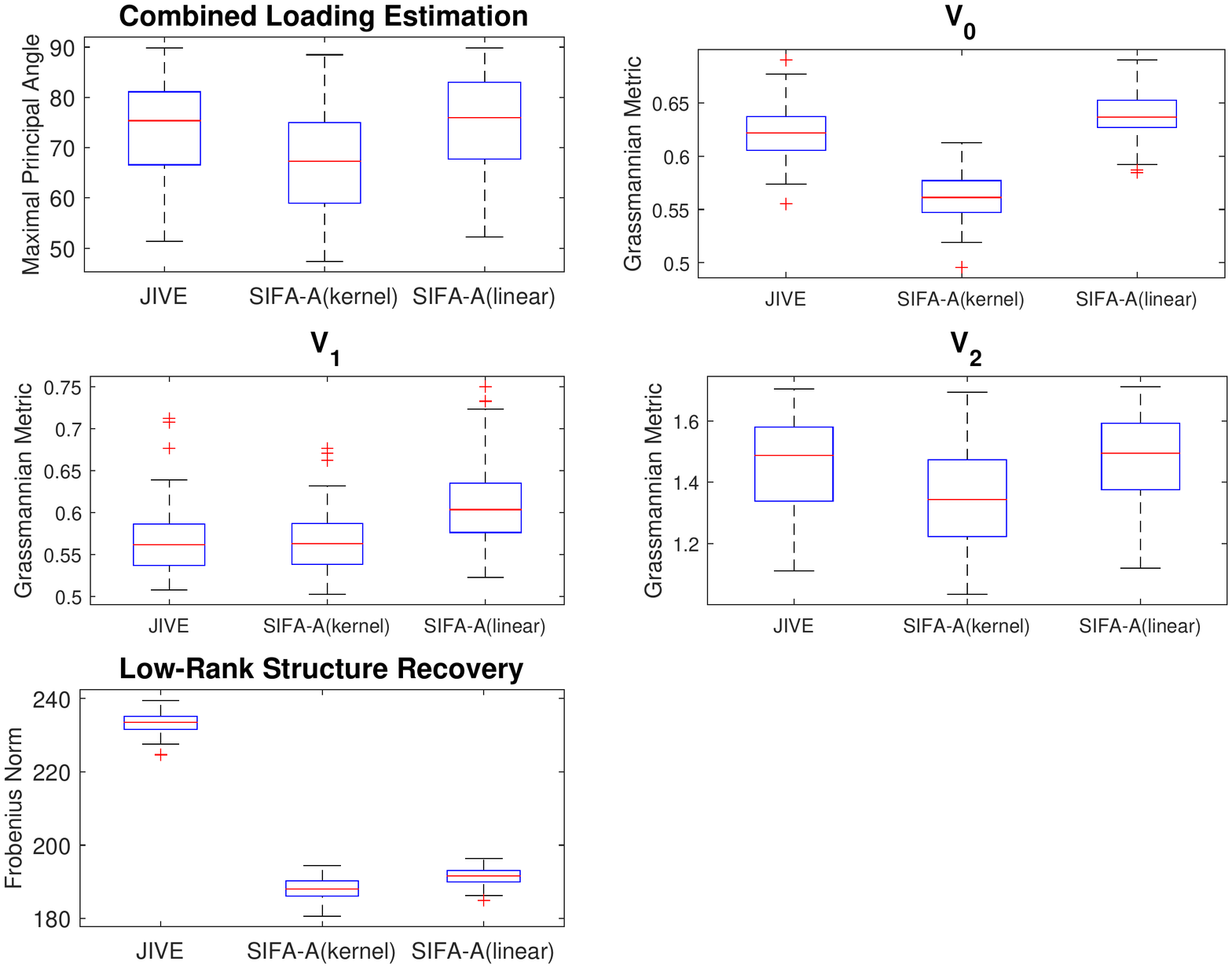}
\end{center}
\newcaption{Simulation study of nonlinear covariate functions. Upper left: the maximal principal angles between the true and estimated loadings under Setting 4; Upper right and middle two: the Grassmannian metric of the differences between the true and estimated loading matrices; Lower left: the Frobenius norm of the differences between the true and estimated low-rank structure. In each figure, from left to right, the box plots correspond to JIVE, SIFA-A with kernel regression, and SIFA-A with linear regression, respectively. The results are based on 100 simulation runs.}
\label{fig:nonlinear}
\end{figure}}

\subsection{Overfitting of Nonparametric Estimation}
\textcolor{black}{We also study the overfitting of nonparametric estimation in SIFA models when the true covariate functions are linear.
In particular, we consider the following simulation setting where the parameters and the ranks are set in the same way as in Setting 3 in the main manuscript, except there is only one covariate ($q=1$).
\bi
\item {\bf Setting 5} (SIFA-B Model with univariate linear relations): The latent factors are generated from $\bU_k=X\bb_k^T+\bF_k$ for $k=0,1,2$, where $X$ represents a univariate covariate and $\bb_k$ is a length-$r_k$ coefficient vector. Other parameters are generated in compliance with the orthogonal conditions.
\ei}

\begin{figure}[!]
\begin{center}
\includegraphics[width=6in]{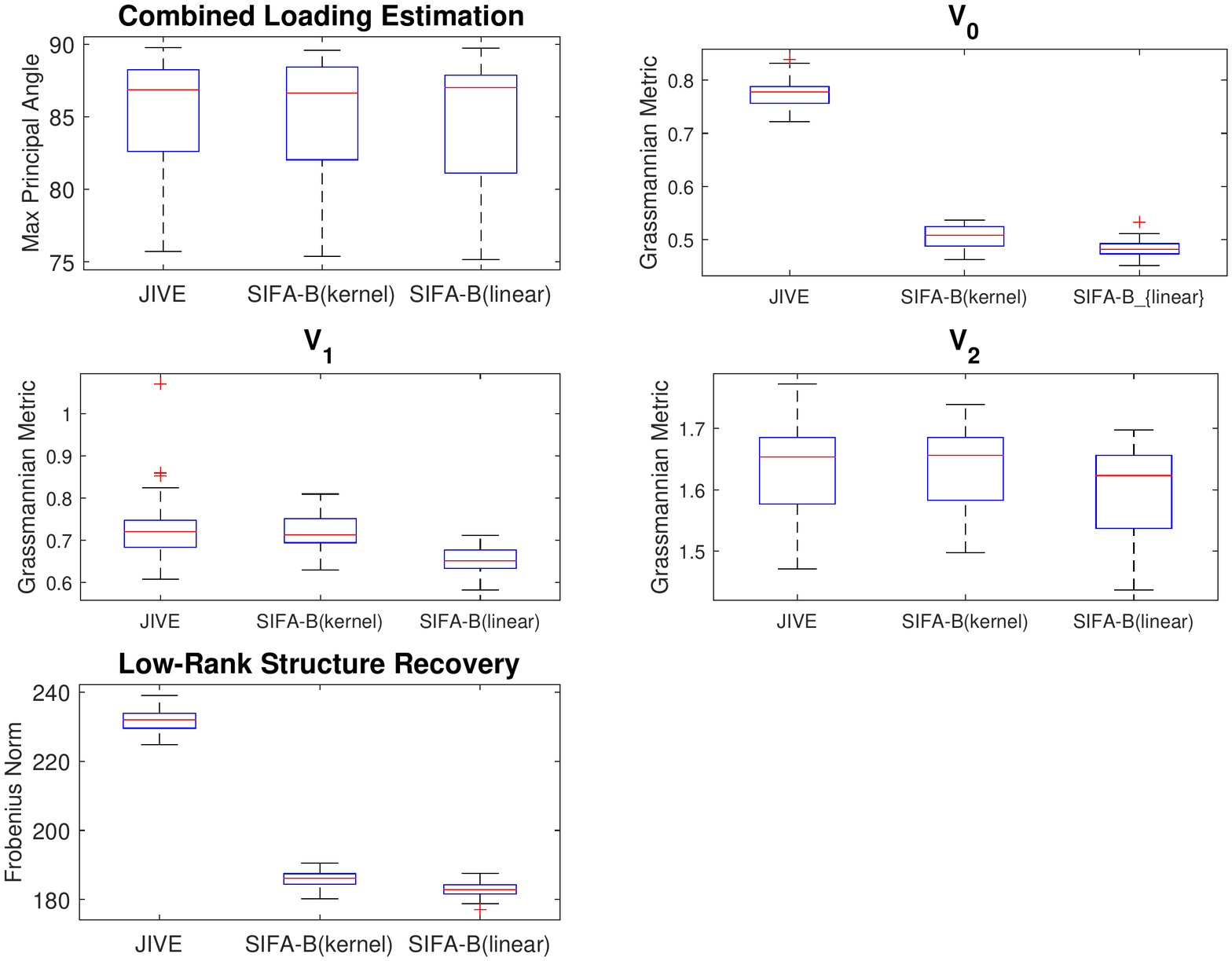}
\end{center}
\newcaption{Simulation study of overfitting of nonparametric estimation. Upper left: the maximal principal angles between the true and estimated loadings under Setting 4; Upper right and middle two: the Grassmannian metric of the differences between the true and estimated loading matrices; Lower left: the Frobenius norm of the differences between the true and estimated low-rank structure. In each figure, from left to right, the box plots correspond to JIVE, SIFA-B with kernel regression, and SIFA-B with linear regression, respectively. The results are based on 100 simulation runs.}
\label{fig:overfit}
\end{figure}

\textcolor{black}{
We focus on SIFA-B methods with nonparametric and linear relations, and the JIVE method.
The comparison results are shown in Figure \ref{fig:overfit}.
SIFA-B fitted with linear regressions (i.e., the true model) has the best estimation accuracy, both for loadings and the low-rank structure.
 SIFA-B fitted with nonparametric regressions has very similar but slightly worse performance, probably due to the overfitting issue.
Both methods significantly outperform the JIVE method.
It indicates the overfitting issue of the nonparametric covariate functions is marginal compared to the exclusion of the auxiliary information (as in JIVE).}

\subsection{Rank Misspecification}
In this section, we provide simulation results for the rank misspecification study.
Data are generated from the Setting 2 (i.e., SIFA-A model).
The true ranks are $r_0=2,r_1=r_2=3$.

\begin{figure}[!]
\begin{center}
\includegraphics[width=6in]{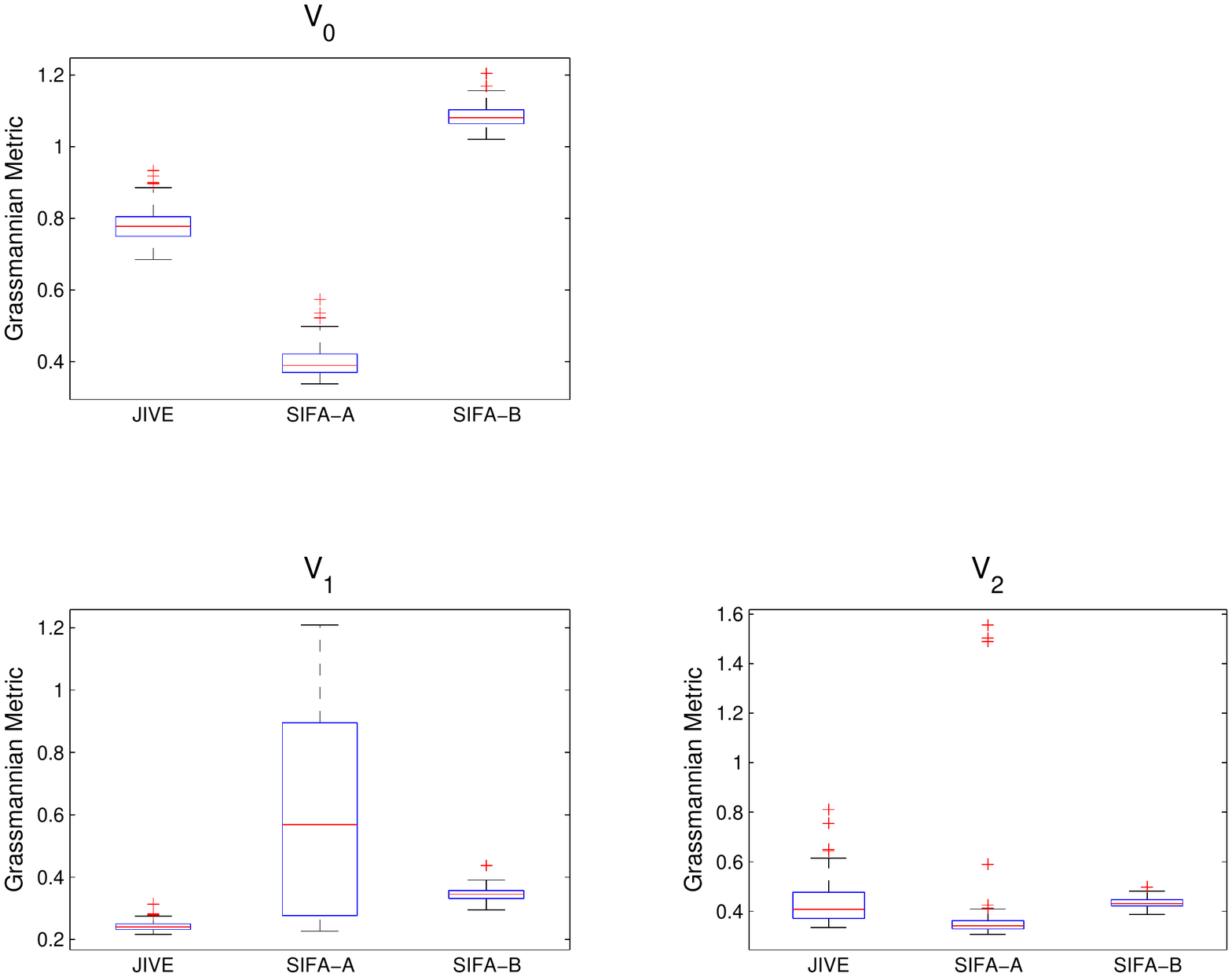}
\end{center}
\newcaption{Simulation study of rank overestimation. The Grassmannian metric of the differences between true and estimated loading matrices under Setting 2 (the estimated loading matrices have overestimated ranks). In each figure, from left to right, the box plots correspond to JIVE, SIFA-A, and SIFA-B respectively. The results are based on 100 simulation runs.}
\label{fig:overV}
\end{figure}

\begin{figure}[!]
\begin{center}
\includegraphics[width=6in]{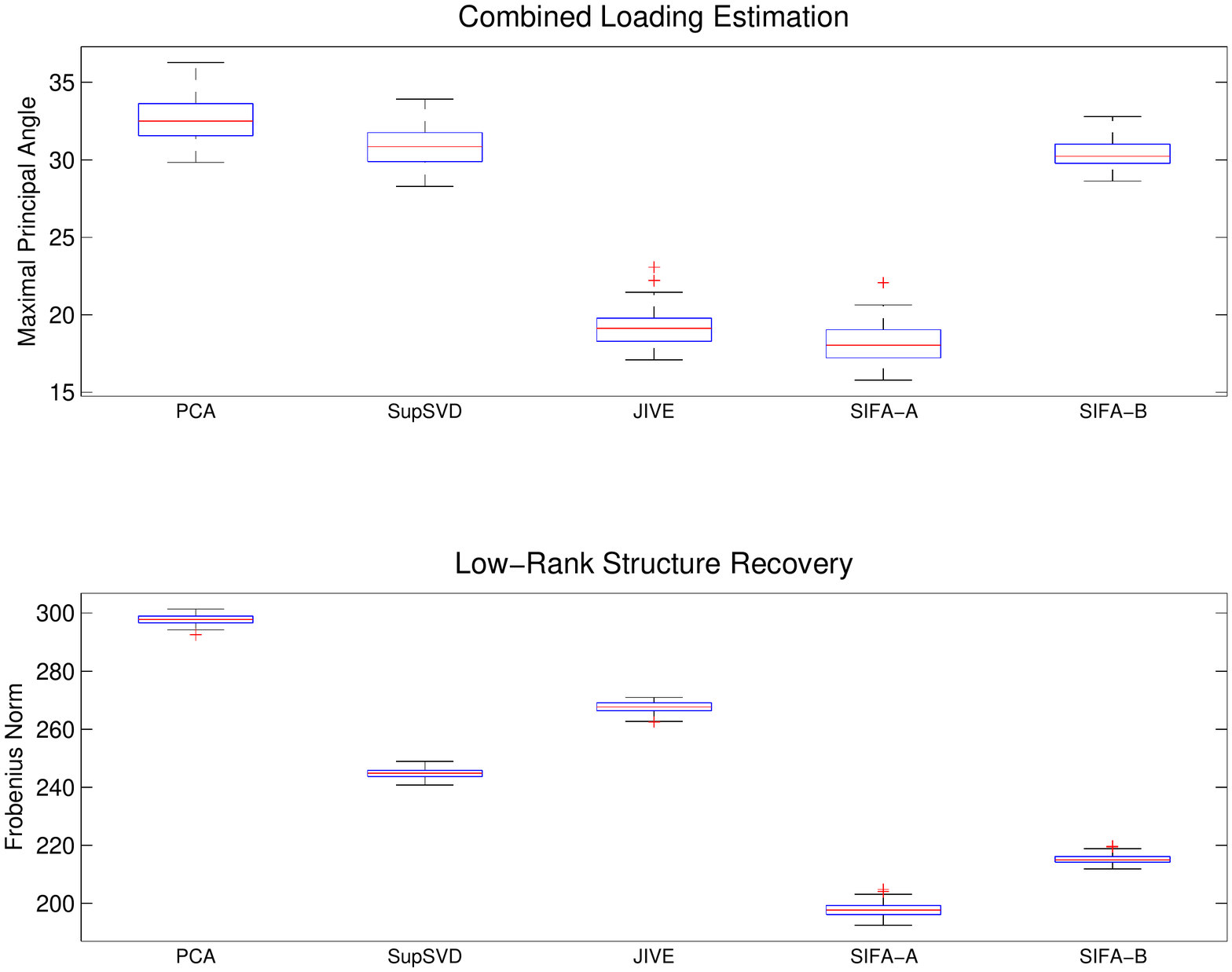}
\end{center}
\newcaption{Simulation study of rank overestimation. Upper: the maximal principal angles between true and estimated loadings under Setting 2 (the estimated loading matrices have overestimated ranks); Lower: the Frobenius norm of the differences between true and estimated low-rank structures. In each figure, from left to right, the box plots correspond to PCA, SupSVD, JIVE, SIFA-A, and SIFA-B respectively. The results are based on 100 simulation runs.}
\label{fig:overAll}
\end{figure}

First, we overestimate every rank by 1. Namely, misspecified ranks $\widehat{r_0}=3,\widehat{r_1}=\widehat{r_2}=4$ are used for fitting JIVE, SIFA-A and SIFA-B, and $\widehat{r_0}+\widehat{r_1}+\widehat{r_2}=11$ is used as the single rank for fitting PCA and SupSVD.
The results are shown in Figure \ref{fig:overV} and Figure \ref{fig:overAll}.
Similar to the simulation results in the main article for Setting 2, SIFA-A outperforms JIVE and SIFA-B in terms of the estimation accuracy of $\bV_0$ and $\bV_2$.
The $\bV_1$ estimate of SIFA-A has a relatively larger variance possibly because one of the individual patterns is captured by the redundant joint rank.
Nevertheless, in the top panel of Figure \ref{fig:overAll}, we can see that SIFA-A provides the best overall estimation accuracy for the combined loadings.
It also has the best low-rank structure recovery accuracy among all methods.

\begin{figure}[htbp]
\begin{center}
\includegraphics[width=6in]{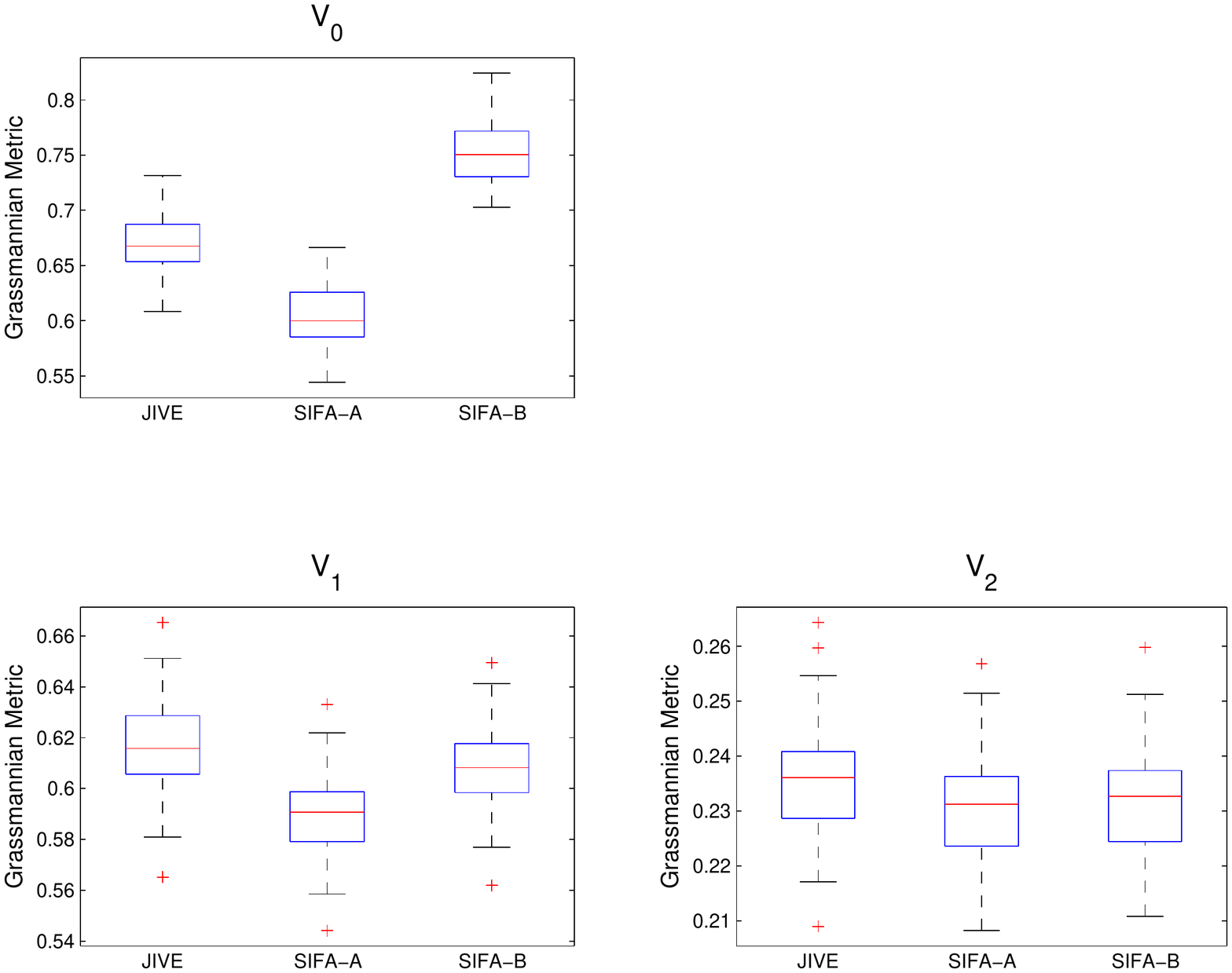}
\end{center}
\newcaption{Simulation study of rank underestimation. The Grassmannian metric of the differences between true and estimated loading matrices under Setting 2 (the estimated loading matrices have underestimated ranks). In each figure, from left to right, the box plots correspond to JIVE, SIFA-A, and SIFA-B respectively. The results are based on 100 simulation runs.}
\label{fig:underV}
\end{figure}

\begin{figure}[htbp]
\begin{center}
\includegraphics[width=6in]{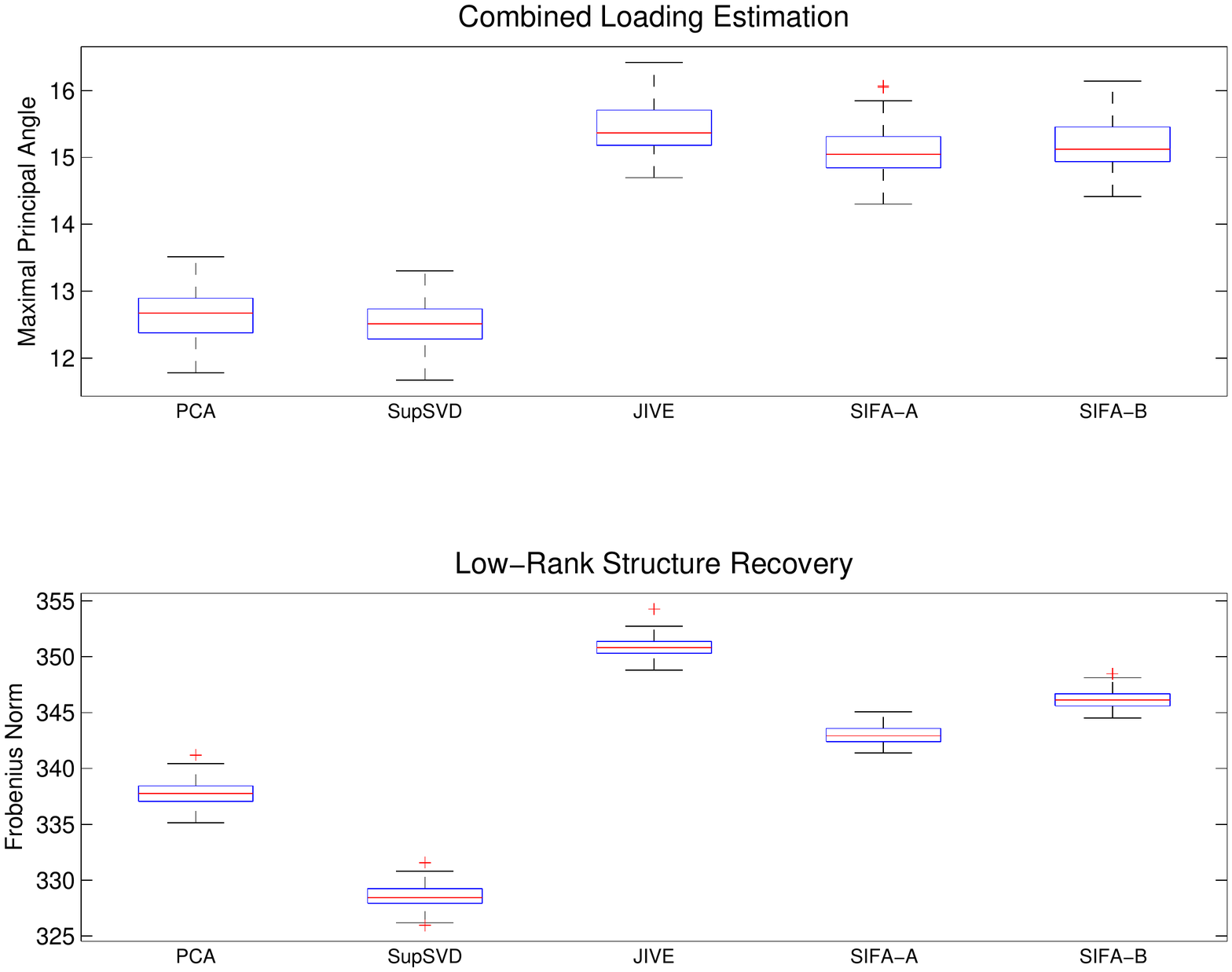}
\end{center}
\newcaption{Simulation study of rank underestimation. Upper: the maximal principal angles between true and estimated loadings under Setting 2 (the estimated loading matrices have underestimated ranks); Lower: the Frobenius norm of the differences between true and estimated low-rank structures. In each figure, from left to right, the box plots correspond to PCA, SupSVD, JIVE, SIFA-A, and SIFA-B respectively. The results are based on 100 simulation runs.}
\label{fig:underAll}
\end{figure}

Then we investigate the effect of underestimated ranks by setting $\widehat{r_0}=1,\widehat{r_1}=\widehat{r_2}=2$.
The simulation results are shown in Figure \ref{fig:underV} and Figure \ref{fig:underAll}.
Again, the results are similar to those obtained from the simulation study under Setting 2 in the main article.
SIFA-A uniformly outperforms JIVE and SIFA-B in separate loading estimation, combined loading estimation and low-rank structure recovery.
Interestingly, we note that in Figure \ref{fig:underAll}, the more sophisticated integrative decomposition methods (i.e., JIVE, SIFA-A and SIFA-B) are worse than the simpler decomposition methods (i.e., PCA and SupSVD for concatenated data) with regard to the combined loading estimation and low-rank structure recovery.
This is mainly because the simpler decomposition methods do not impose any rank restriction for different patterns.
They have the freedom to allocate ranks optimally to capture the underlying structure with the largest variation.
When all the ranks are underestimated, the higher flexibility of rank allocation leads to the better overall performance.
Nonetheless, the integrative decomposition methods (especially the proposed SIFA methods) have the unique capacity of distinguishing joint and individual patterns, which potentially enhance interpretation and facilitate further statistical analyses.
Therefore, they are still deemed very useful in practice even when the ranks are misspecified.

\subsection{Non-Gaussian Noise}
\begin{figure}[htbp]
\begin{center}
\includegraphics[width=6in,height=4in]{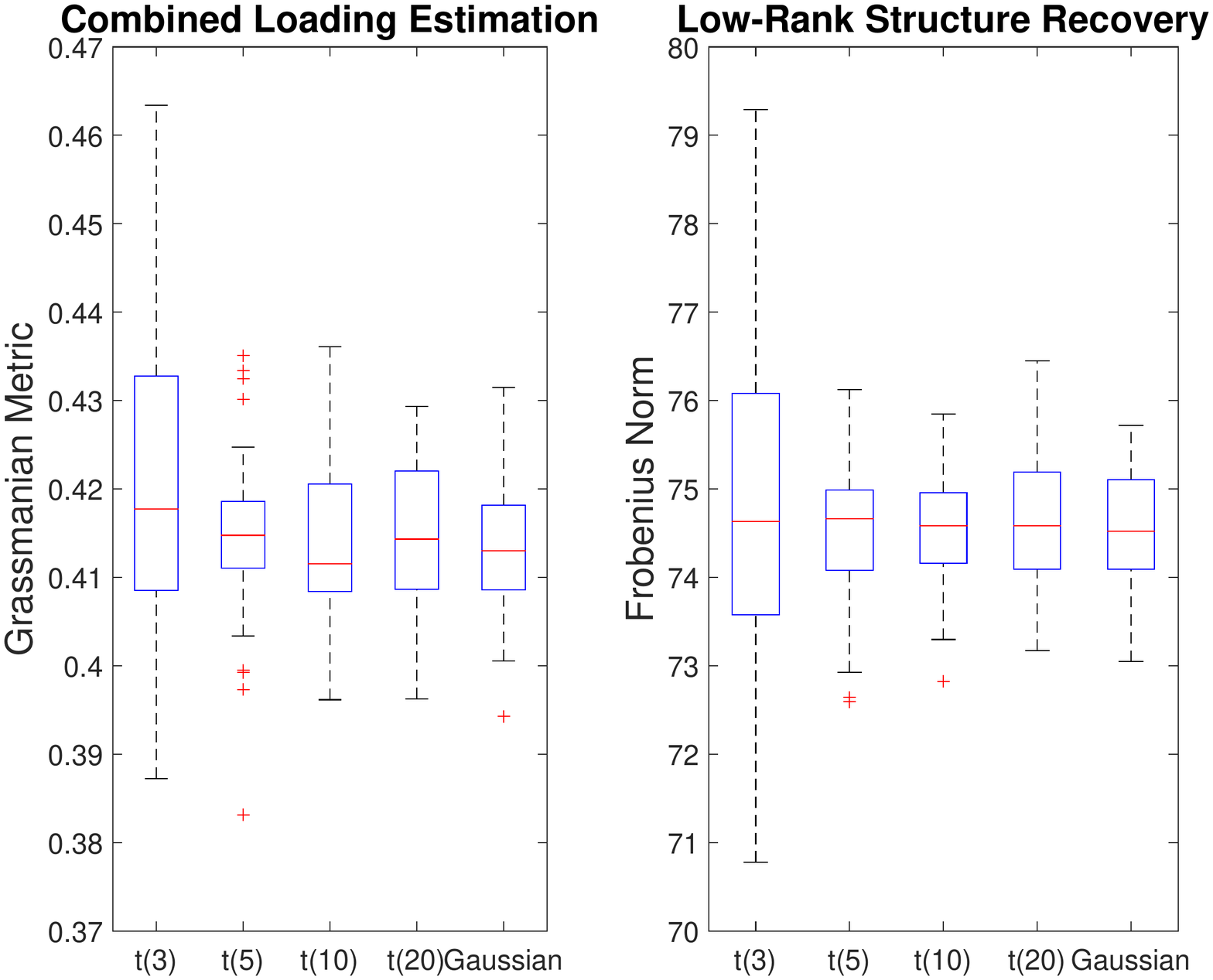}
\end{center}
\newcaption{Simulation study of non-Gaussian noise. Left: the Grassmanian metric between the true and the estimated loadings; Right: the Frobenius norm of the differences between the true and the estimated low-rank structures. In each panel, from left to right, the box plots correspond to the noise distributions $t(3)/\sqrt{3},\ t(5)/\sqrt{5/3} ,\ t(10)/\sqrt{5/4} ,\ t(20)/\sqrt{10/9}$ and $\mathcal{N}(0,1)$.}
\label{fig:nongaussian}
\end{figure}

\textcolor{black}{In this section, we investigate the influence of the non-Gaussian noise on model fitting.
We exploit the simulation Setting 3 in the main manuscript, except for the measurement error matrices $\bE_1$ and $\bE_2$.
We assume the noise variables in $\bE_1$ and $\bE_2$ are i.i.d.\ from the same distribution, and consider 5 different distributions: the standard Gaussian distribution, and the $t$ distributions with the degree of freedom 3, 5, 10, and 20, respectively.
In particular, for each $t$ distribution, we multiply the error matrices with a constant (i.e., $\sqrt{(df-2)/df}$) so that the random variables have unit variance.
Then we fit the SIFA-B model, and evaluate the loading estimation and low-rank structure recovery accuracy under different settings.
The results are presented in Figure \ref{fig:nongaussian}.
We remark that the estimation is not too much affected by the violation of the Gaussian assumption.
When the noise distribution is very heavy-tailed (i.e., $t(3)/\sqrt{3}$), the variation of the estimation over different simulation runs increases.
This is possibly due to the effect of the outliers in the noise matrix.
All in all, the method is very robust when the noise is non-Gaussian.}

\subsection{Rescaling of Different Data Sets}
\textcolor{black}{
Statistical decisions or actions based on data should not be affected by simple transformation. Here we first show that the SIFA models under the general conditions are equivariant under individual scaling of data sets, and extend the results to the case of highly-imbalanced dimensions of data sets.}
Suppose we have a sample of size $n$ for $K=2$ data sets from a SIFA model satisfying the orthogonal conditions:
\begin{align}
\label{eq:simul_model_1}
\bY_1 &= (\bdf_0(\bX) + \bF_0) \bV_{0,1}^T + (\bdf_1(\bX) + \bF_1)\bV_1^T + \bE_1, \\
\bY_2 &= (\bdf_0(\bX) + \bF_0) \bV_{0,2}^T + (\bdf_2(\bX) + \bF_2)\bV_2^T + \bE_2. \label{eq:simul_model_2}
\end{align}
If the measurement unit of  $\bY_1$ has changed so that all entries of $\bY_1$ is scaled by some $s > 0$, then the combined model for the scaled data becomes
\begin{align}
\label{eq:simul_model_s}
[s\bY_1; \bY_2] = c_s( \bdf_0(\bX) + \bF_0) \bV_{0(s)}^T + [s \bdf_1(\bX) + s\bF_1 ; \bdf_2(\bX) + \bF_2 ] \mbox{blkdiag}(\bV_1^T, \bV_2^T) + [s\bE_1; \bE_2]
\end{align}
 where $c_s =  s^2 /2 + 1/2$ and  $\bV_{0(s)}^T = [s\bV_{0,1}^T ;\bV_{0,2}^T]/c_s$. This \emph{scaled} model does not  satisfy the orthogonal conditions any more, but it still satisfies the general conditions. Since for any value of $s$, the model consists of the same direction vectors $\bV_{0,1}, \bV_{0,2}, \bV_{1}$ and $\bV_{2}$ and the interpretation of the scores remains the same for any $s$, the SIFA model under the general conditions is equivariant under arbitrary  scaling of individual data sets.

\begin{figure}[htbp]
\begin{center}
\includegraphics[width=6in]{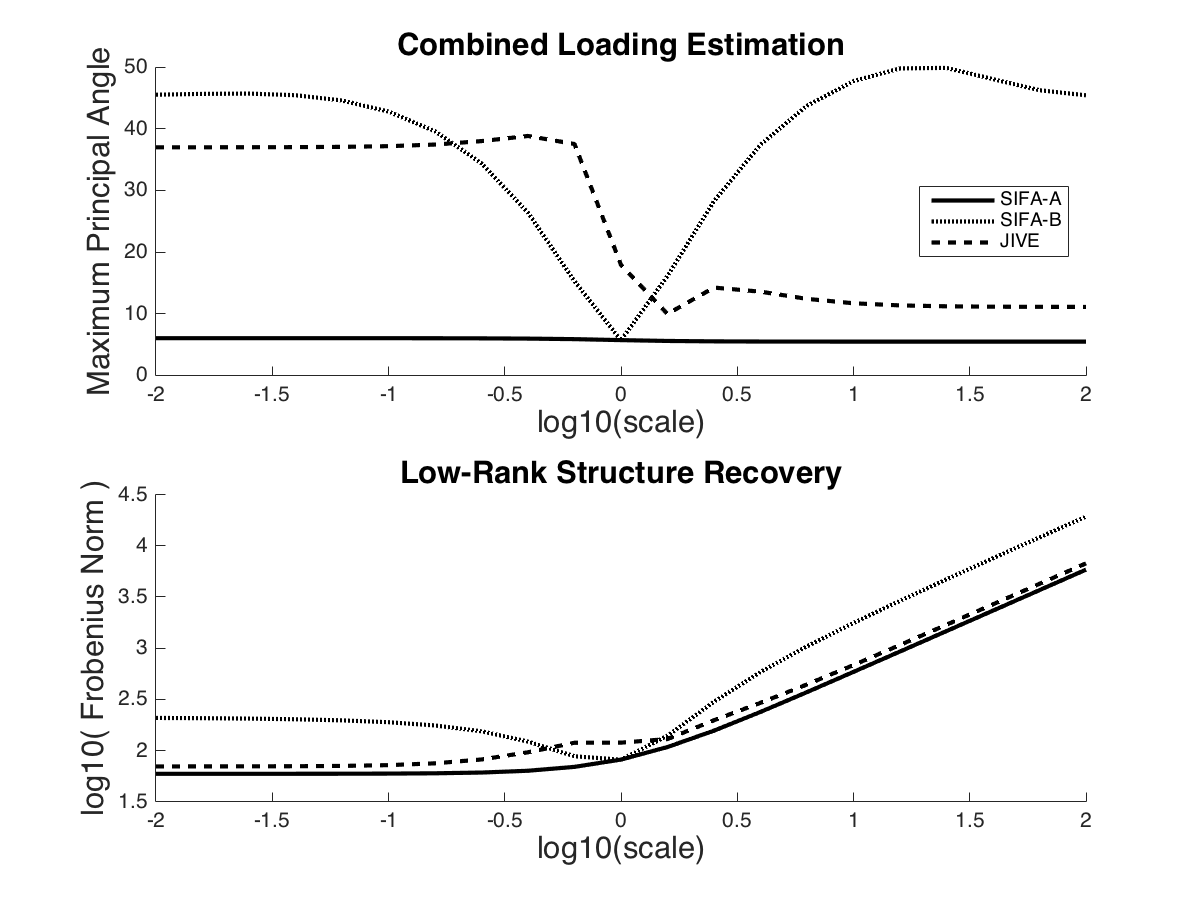}
\end{center}
\newcaption{Simulation study for data scaling. Top: the maximal principal angles between the true ($[\bV_{0(s)}, \mbox{blkdiag}(\bV_1^T, \bV_2^T)]$) and estimated loadings; Bottom: the Frobenius norm of the differences between the true and estimated low-rank structure. The true loading and low-rank structure for each scale value, $s$, are from  (\ref{eq:simul_model_s}).
}
\label{fig:SIFA-Scale-Equivariance-1}
\end{figure}

\begin{figure}[htbp]
\begin{center}
\includegraphics[width=6in]{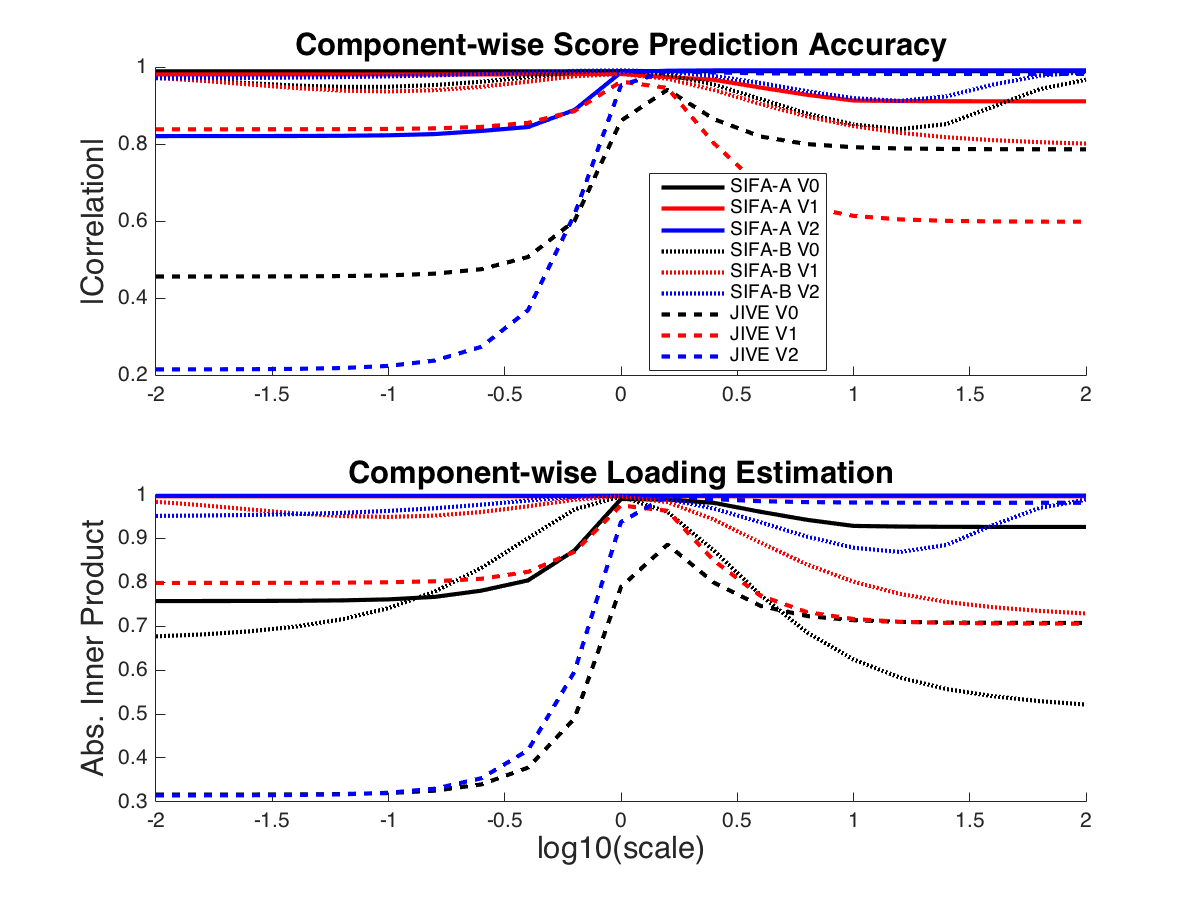}
\end{center}
\newcaption{Simulation study for data scaling. Top: the absolute correlation coefficients between the true $\bdf_i(\bX) + \bF_i$ and its predictions ($i=0,1,2$); Bottom: The inner product between the true and estimated loadings.  The true loading and low-rank structure for each scale value, $s$, are from  (\ref{eq:simul_model_s}).  }
\label{fig:SIFA-Scale-Equivariance-2}
\end{figure}

\textcolor{black}{
While the SIFA model under the general conditions is equivariant to individual scaling, the likelihood-based estimators can be sensitive to relative sizes of data sets. Nevertheless, we have observed that the SIFA estimates under general conditions, SIFA-A estimates, are \textit{nearly equivariant}.
We demonstrate the near-equivariance of SIFA-A estimates for a simulated data set, with a comparison to SIFA-B estimates and JIVE estimates. For this experiment, we have generated a sample from the SIFA model (\ref{eq:simul_model_1})-(\ref{eq:simul_model_2}) satisfying the orthogonal conditions. For simplicity, we used rank 1 for each of joint and individual variations. The first data set $\bY_1$ is then individually rescaled by $s$ ranging from 0.01 to 100; that is, in one extreme, the Frobenius norm of $\bY_1$  is approximately 100 times larger than that of $\bY_2$ (and vice versa). For this example, we check that the loading estimates of SIFA-A are nearly equivariant under individual scaling, as shown in Figures  \ref{fig:SIFA-Scale-Equivariance-1} and \ref{fig:SIFA-Scale-Equivariance-2}. On the other hand, the estimated loadings of SIFA-B and JIVE are quite dependent to the scale, and the estimation accuracy of those is worse than that of SIFA-A. Note that for any $s \neq 0$, the true model violates the orthogonal conditions of SIFA-B. The low-rank structure recovery of SIFA-A is also among the best across various scales (see Figure \ref{fig:SIFA-Scale-Equivariance-1}).}

\textcolor{black}{
We further point out that the predicted scores from each rescaled data sets are highly correlated to each other. Moreover, as shown in Figure \ref{fig:SIFA-Scale-Equivariance-2}, the predictions of SIFA-A are highly correlated with the true scores with correlation coefficients $>0.8$ for all values of $s$ considered. Thus, any subsequent analysis using the score predictions of SIFA-A are invariant to individual scaling of data sets. On the other hand, JIVE does not possess such equivariance property and the correlation coefficients to their predictions are as low as 0.2. In this and many other examples, we have observed that SIFA-B score predictions were as good as those of SIFA-A. Although SIFA-B loading estimates are not as accurate as those of SIFA-A, they are at most about 50 degrees away from the truth. Heuristically, the projections of data points along the true $\mathbf{v}$ to $\hat{\mathbf{v}}$ preserve the order and relative magnitudes of original scores, if they are not orthogonal, which may explain the high score prediction accuracy of SIFA-B. Overall, in terms of score prediction, SIFA-A and SIFA-B both have shown the invariance to individual scaling.}

\textcolor{black}{We also investigate the SIFA model fitting when the the dimensions of the individual data sets are highly imbalanced. We remark that the SIFA-B model is robust against imbalanced dimensions, because of the orthogonal and equal norm identifiability conditions. For the SIFA-A model, it is challenging to derive an argument similar to the equivariance under the rescaling of data. However, we observed the near-equivariance under two scenarios of increasing dimensions (of the first data set), while the second data set is kept fixed. The first scenario corresponds to the case where variables in the increasing dimension has factors highly correlated to existing factors, while the second scenario is the case where the added variables are pure noises. In each of both cases, we artificially added variables to make dimensions imbalanced (up to $(p_1, p_2) = (5000, 100)$). Notably, the proportion of $\|\hat{V}_{0,1}\|^2$ to $\|\hat{V}_{0,1}\|^2 + \|\hat{V}_{0,2}\|^2$ increases as $p_1$ increases (but at a very slow rate). In the meantime, the loading estimates of the SIFA-A model remain to be significantly better than the JIVE estimates (graphical results are omitted, but similar to Figures S10-S11) as $p_1$ increases. We remark that the first scenario of increasing dimensions produces a similar log-likelihood function as that from scaling the first data set by $s = p_1 / p_2$. Thus, the near-equivariance is indeed expected. }

\subsection{Scalability of SIFA Algorithms}
\begin{figure}[htbp]
\begin{center}
\includegraphics[width=4in]{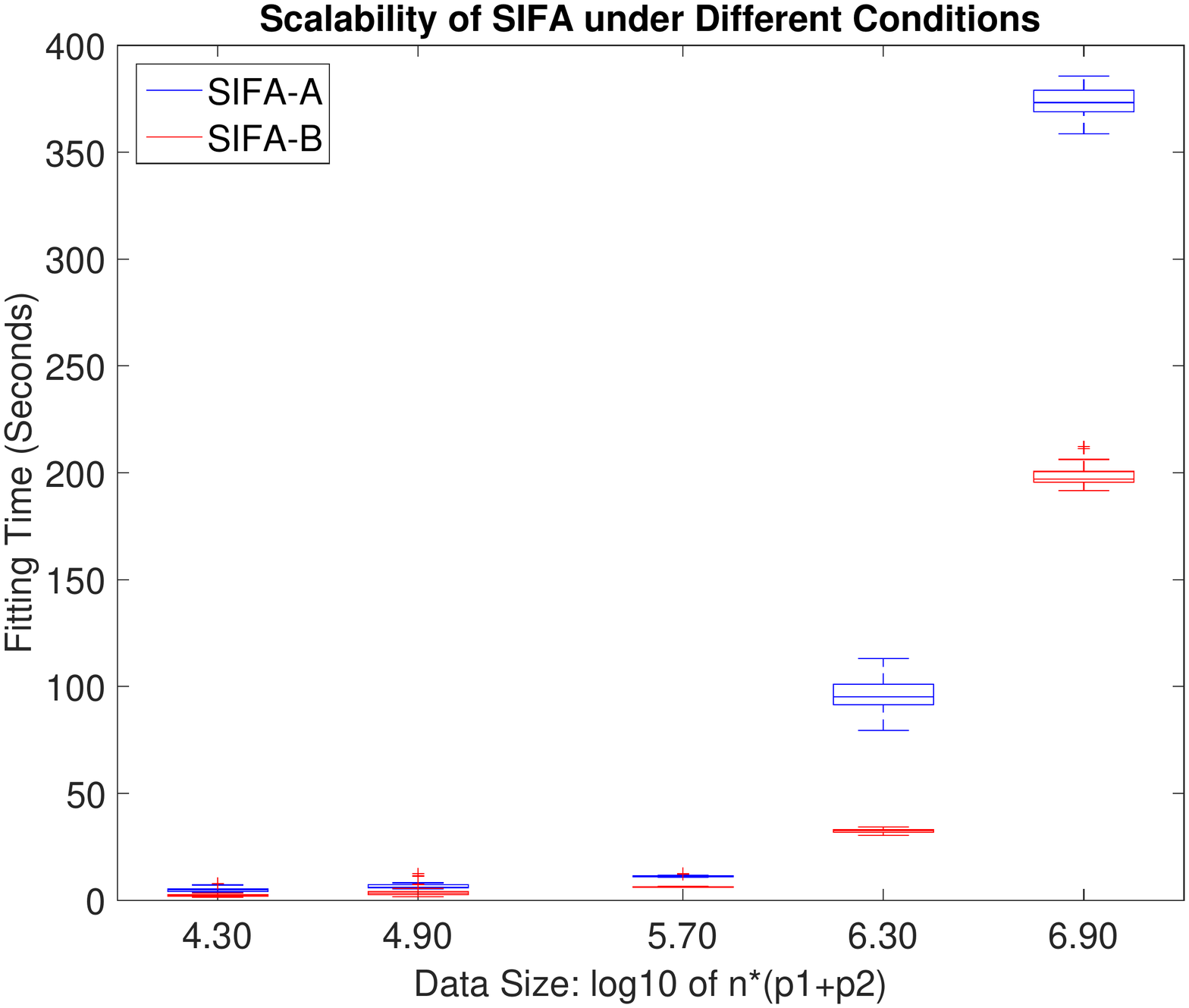}
\end{center}
\newcaption{Scalability of the SIFA model fitting algorithms. From left to right, we consider simulation settings with the sample size-dimension tuples $(n,p_1,p_2,q)\in\{(100,100,100,10),\\ (200,200,200,20), (500,500,500,50),(1000,1000,1000,100), (2000,2000,2000,200)\}$. The blue boxes correspond to the fitting times of SIFA-A while the red boxes correspond to the fitting times of SIFA-B. The x-axis is on the $\log_{10}$ scale of the total number of entries in the primary data sets.
The result is based on 30 simulation runs in each setting.}
\label{fig:time}
\end{figure}

\textcolor{black}{
We also investigate the scalability of the proposed Algorithms \ref{alg1} and \ref{alg2} for model fitting.
We exploit the simulation Setting 3 in the main manuscript and consider a range of dimensions to generate data. In particular, we consider 5 settings with the sample size $n$, the dimensions $(p_1,p_2)$, and the number of covariates $q$ being: $(n,p_1,p_2,q)=(100,100,100,10), (200,200,200,20), (500,500,500,50),\\(1000,1000,1000,100), (2000,2000,2000,200)$.
For each setting, we conduct 30 simulation runs.
In each simulation run, we fit SIFA-A and SIFA-B to the data.
The computing time on a standard desktop is summarized in Figure \ref{fig:time}.
We remark that both algorithms are highly efficient.
Even for thousands of samples and dimensions, it only takes a few minutes to fit a SIFA model on a desktop.
The SIFA-B algorithm is more computationally efficient than the SIFA-A algorithm since the M step has explicit solutions.
The directions for further improvement include parallel computing and distributed computing.
}

\section{GTEx Data Analysis}\label{suppsec:GTEx}
\textcolor{black}{In this section, we provide additional details on the data preprocessing and rank estimation of the GTEx example.}

\subsection{Data Preprocessing}
\textcolor{black}{
We focus on three tissues with the most samples in the GTEx data, i.e., muscle ($n=361$), blood ($n=338$), and skin ($n=302$).
There are 204 common samples across 3 tissues.
The p53 signaling pathway contains around 200 genes (the gene names can be found on the GSEA website \url{http://software.broadinstitute.org/gsea/msigdb/cards/HALLMARK_P53_PATHWAY.html}).
After removing the non-expressed genes according to the GTEx preprocessing criteria, we end up with 191 genes in each tissue.
We normalize the expression level of each gene in each tissue through an inverse normal transformation as in \cite{ardlie2015genotype}.
Consequently, each gene follows a normal distribution with mean zero and unit variance.
We denote the three preprocessed expression matrices as $\bY_1,\bY_2,\bY_3$, which are the primary input data for the SIFA method.}

\begin{figure}[H]
\begin{center}
\includegraphics[width=4.5in]{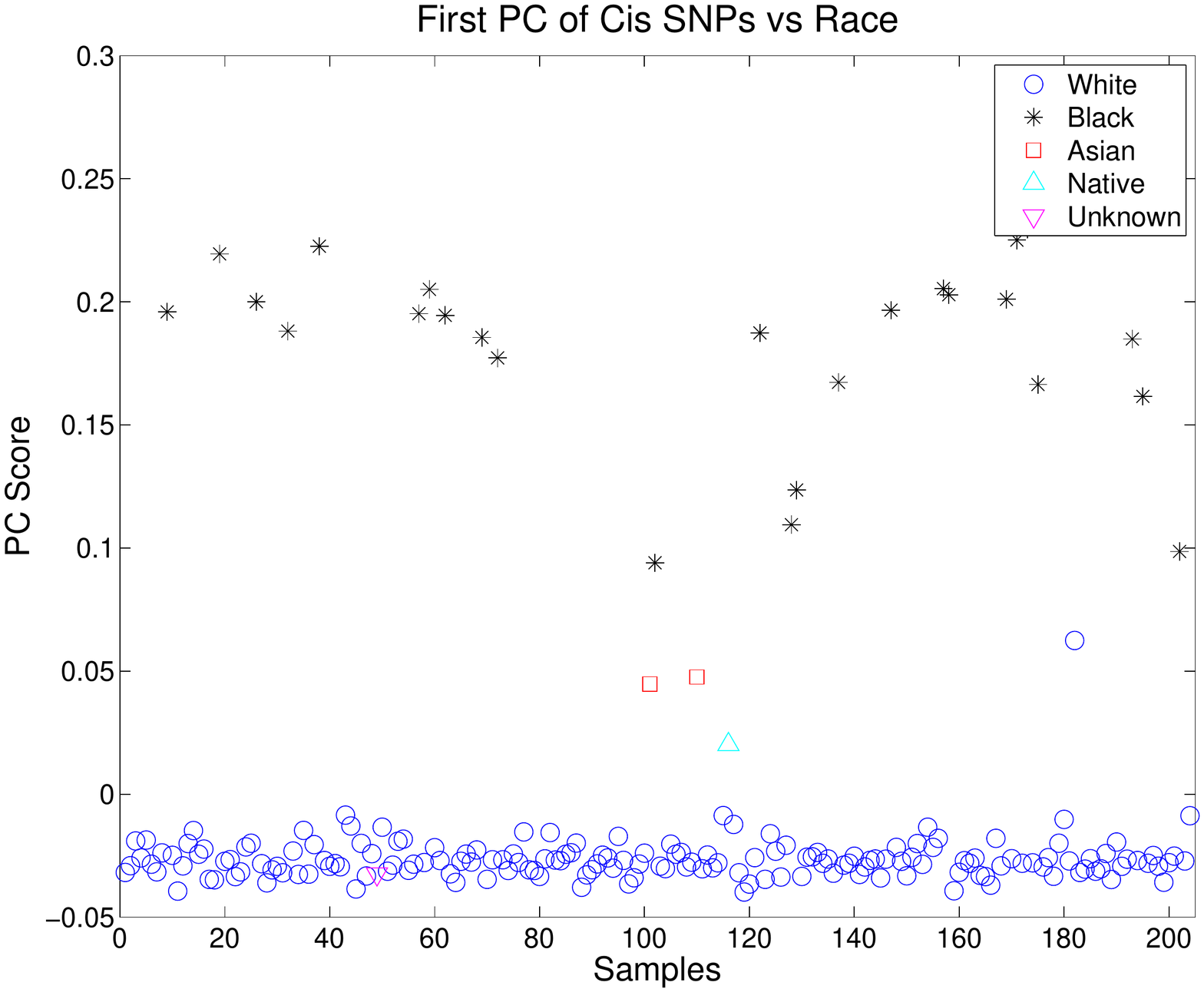}
\end{center}
\newcaption{GTEx Example: the first PC score of the cis SNP data and the race information. The majority of the population is white (about $85\%$); the second largest group is African American (about $14\%$); some individuals are Asian or native American, or have unknown information.}
\label{fig:PC1}
\end{figure}

\textcolor{black}{
For each of the 204 common samples, we also have the sex information, the genotyping array platform index, and the individual's genotype information.
The genotype data contain the mutation status of 6,856,774 single nucleotide polymorphisms (SNPs).
We particularly restrict our scope to the {\em cis} SNPs of the genes in the p53 pathway, i.e., SNPs that lie within 1 megabase pair (1Mbp) of the transcription start site of a gene.
Moreover, we eliminate those SNPs with minor allele frequency smaller than $10\%$ across samples.
As a result, we obtain 639,965 SNPs.
The dimension of the genotype data far exceeds the sample size.
To further reduce the dimension of the genotype data, we apply PCA to the data and use the top 30 PC score vectors to represent the genotype data.
We note that the first PC score vector typically captures the ancestry information (see Figure \ref{fig:PC1}) and the other PC scores capture the majority of variation in the SNP data.
We column-center the two binary variables and the 30 PC score vectors and treat them as the covariates, denoted by $\bX$, for the SIFA method.}

\textcolor{black}{Because our primary goal is to identify gene expression patterns across tissues and quantify the regulatory effect of SNP mutations on gene expressions, we treat the genotype data as covariates in this example.
We care more about the information contained in the genotype data rather than the specific mutation status of each SNP.
It is thus plausible to use the top principal components of the genotype data as a surrogate of the original ultra-high dimensional data in the SIFA model.
However,  if one is interested in integrating the genotype data and the gene expression data, raw data should be used.
In that scenario, both data sets are  primary data sets.
we remark that one caveat of using the raw genotype data in that case is that the data may only take binary values or ternary values.
It does not satisfy the Gaussian assumption of the SIFA model.
Extending the model to accommodate non-Gaussian data calls for more investigation.
  }

\subsection{Rank Estimation}
\begin{figure}[htbp]
\begin{center}
\includegraphics[width=6in]{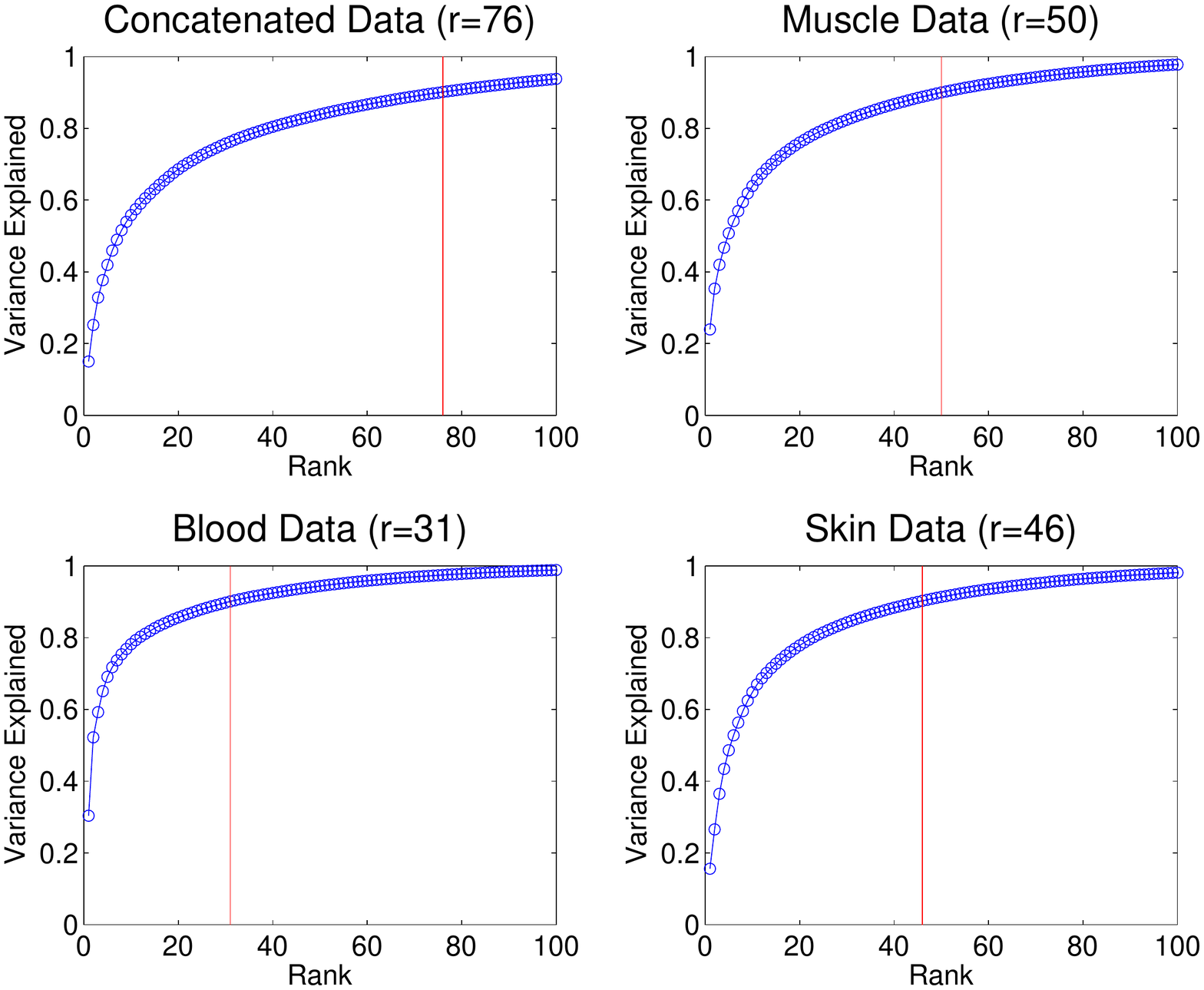}
\end{center}
\caption{GTEx Example: The variance explained by top eigenvalues in different data sets (top left: concatenated data; top right: muscle data; bottom left: blood data; bottom right: skin data). The vertical red line in each plot indicates the smallest rank at which the variance explained exceeds $90\%$ of the total variance.}
\label{fig:rank}
\end{figure}

\textcolor{black}{In order to apply SIFA, we need to estimate the ranks for the joint structure and individual structures first.
By visually inspecting the scree plot of the singular values of each data set, we do not observe any clear elbow point.
Therefore, we exploit a variance explained criterion to estimate the signal ranks of different data sets in the first step of the two-step procedure described in the main manuscript.
By setting the threshold to be $90\%$, we estimate of the intrinsic ranks for different data matrices to be $r^\star_{total}=76, r_1^{\star}=50, r_2^\star=31,r_3^\star=46$.
The actual proportions of variance explained against the ranks in different data sets are shown in Figure \ref{fig:rank}.
Then by solving the equation system, in the second step, we obtain the joint and individual ranks for SIFA as $r_0=26$, $r_1=24$, $r_2=5$, and $r_3=20$.
Since the ranks are all very large and each rank only explains a moderate amount of variation in each data set, it is reasonable to perceive that the result is insensitive to minor changes of the ranks.
Thus we omit the refining step through LCV.
Note that the individual rank for blood  ($r_2=5$) is much smaller than that for muscle or skin.
To some extent, this is consistent with the previous GTEx finding that blood is an outlier tissue \citep{ardlie2015genotype}.}

\subsection{Model Fitting}
\textcolor{black}{
We  fit a SIFA-B model to the data with linear relations between the covariates and the latent factors, for the following reasons.
First of all, according to the preprocessing procedure, different primary data matrices have comparable scales (actually identical in this data example).
It is reasonable to assume that different data contribute equally to the joint factors, and thus the norm constraint in Condition B1 holds.
Second, the high dimensionality of the gene expression data justifies the orthogonality constraints in Condition B1 and B2.
Third, the linear relations between the covariates and the factors lead to high interpretability.
The orthogonality between different components and the linear models make it very easy to decompose the total variation of the primary data into unrelated parts.
Last but not least, from a computational perspective, the SIFA-B model with linear relations is extremely fast to fit (the algorithm converges with high precision within 1 minute on a standard desktop computer).
All in all, we fit an SIFA-B model with linear relations to the data.}

\section{Berkeley Growth Study}\label{suppsec:growth}
\begin{figure}[!]
\begin{center}
\includegraphics[width=6in]{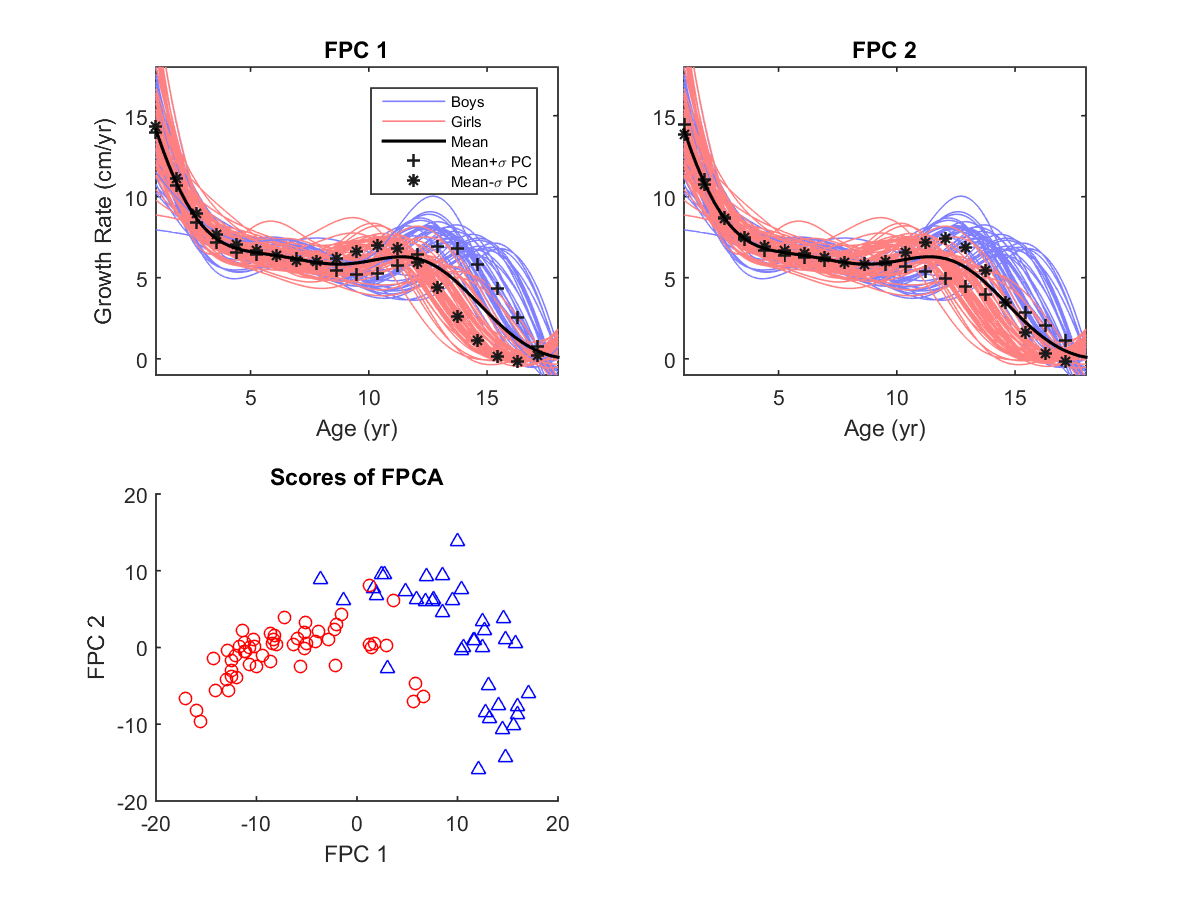}
\end{center}
\newcaption{Berkeley Growth Rate Example: FPCA results. Top panels: the raw growth rate curves overlaid with the mean curve and the mean curve plus/minus one unit of the standard deviation of FPC (FPC1 on the left, and FPC2 on the right). Bottom panel: the scatter plot of FPC scores. Boys (blue triangles) and girls (red circles) are marked differently.}
\label{fig:FPCA}
\end{figure}

In this section, we apply SIFA to the Berkeley Growth Study data set \citep{ramsay2002applied}.
The growth data consist of the height measurements of 39 boys and 54 girls from age 1 to 18. We particularly focus on the growth rate curves derived from the height measurements.
As a preliminary analysis, we apply the conventional functional PCA (FPCA) \citep{ramsay2002applied} to the data set.
The first two principal loadings and scores are shown in Figure \ref{fig:FPCA}.
The FPCA provides a low-rank representation of the data, but the quality of the dimension reduction is questionable.
From the loading plots, we can see that the phase information and the amplitude information of the growth rates are highly confounded.
The scatter plot of the scores shows there is a strong sex effect which is not incorporated into the FPCA.
In addition, the two score vectors are highly associated with each other in a nonlinear way.
All in all, the FPCA does not fully capture the underlying structure of the data, and may lead to misleading interpretation of the pediatric growth patterns.

To provide more effective dimension reduction and to better understand the population growth patterns, we decouple the original growth rate data into two related data sets, representing the ``amplitude'' and ``phase'' of growth respectively. This decomposition exploits a time-warping alignment of the observed functions to a common mean. The time-warping functions become the phase data set $\mathbf{Y}_2$, and the resulting aligned growth rate functions form the amplitude data set $\mathbf{Y}_1$. See \cite{lee2016combined} for a detailed exposition on the decomposition.
Then we apply SIFA to the dual data sets $(\mathbf{Y}_1,\mathbf{Y}_2)$ jointly, and treat sex $X$ as the univariate covariate.
In particular, we fit the model under the general conditions, and exploit linear functions to capture the relations between sex and the latent factors.

\begin{figure}[htbp]
\begin{center}
\includegraphics[width=6in]{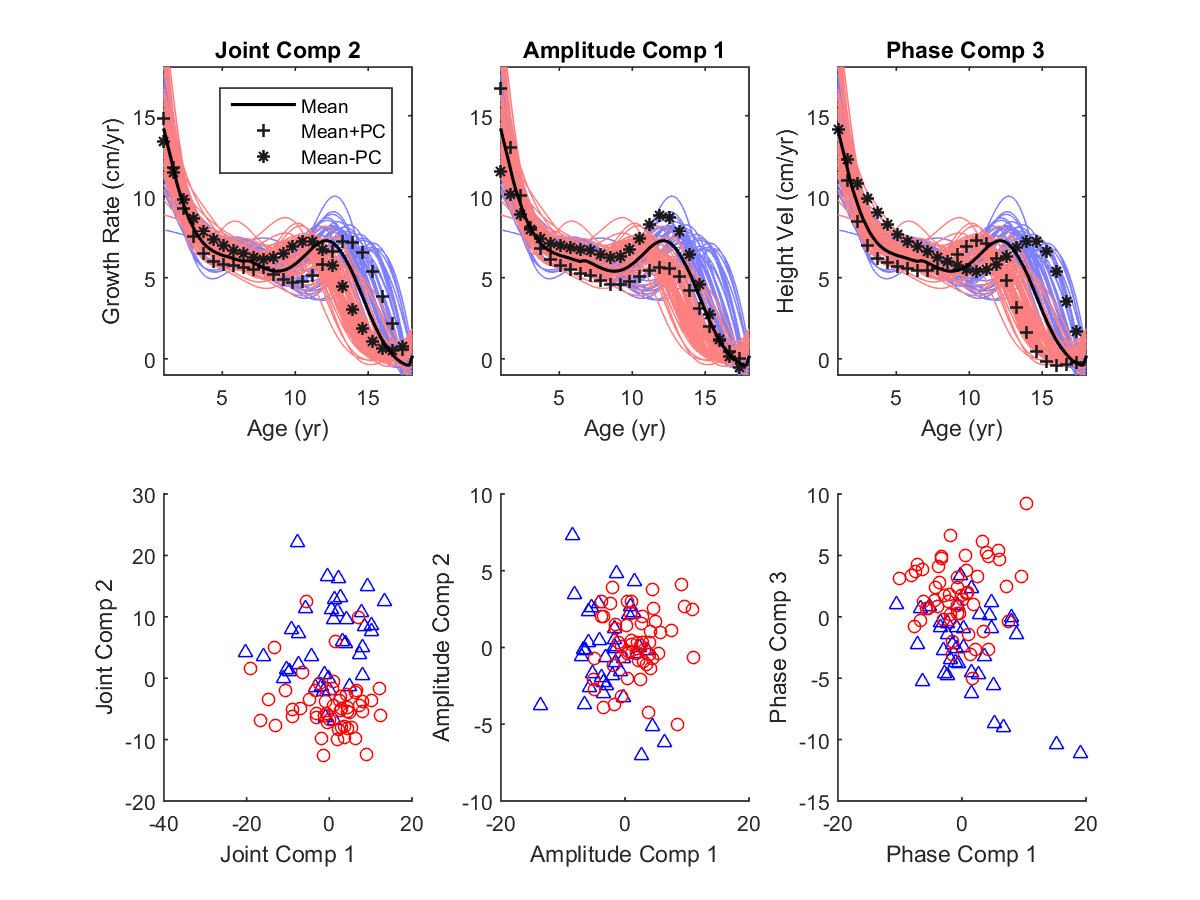}
\end{center}
\newcaption{Berkeley Growth Rate Example: SIFA results. Top panels: the raw growth rate curves overlaid with the mean curve and the mean curve plus/minus one unit of the second joint loading (left), the first amplitude loading (middle), and third phase loading (right). Bottom panels: scatter plots of the first and second joint factors (left), the first and second amplitude factors (middle), and the first and third phase factors (right). Boys (blue triangles) and girls (red circles) are marked differently. Here the joint and individual components are selected based on how much they capture the sex effect.}
\label{fig:SIPCA}
\end{figure}

By fixing the threshold for variance explained to be $90\%$ and following the two-step procedure for rank estimation discussed in the main article, we obtain $(r_0, r_1,r_2) = (2,2,4)$. We model each data set as
\bes
\mathbf{Y}_k = \sum_{j=1}^{r_0} (\beta_{0,j}X + F_{0,j})V_{0,j}^T + \sum_{j=1}^{r_k} (\beta_{k,j}X + F_{k,j})V_{k,j}^T + \bE_k,\ k = 1, 2.
\ees
When estimating the coefficient parameter $\beta_{k,j}$, we use a LASSO estimator \citep{tibshirani1996regression} to incorporate variable selection.
Table~\ref{tab:SIPCA-A-Growth1} lists the estimated coefficients $\beta_{k,j}$, and the standard deviations  for the random signals in $\bF_0,\bF_1,\bF_2$ and the random noise in $\bE_1,\bE_2$.
For better interpretation, the estimated loadings are shown in the original functional space in Figure~\ref{fig:SIPCA}, in a way similar to depicting the FPCA loadings. In addition, the conditional expectations of the latent factors are plotted against each other to visually understand the quality of the dimension reduction.

\begin{table}[h]
\caption{Berkeley Growth Rate Example: Estimated parameters of the SIFA model under the general conditions. The estimated coefficients for the sex effect are denoted by $\widehat{\beta}$ (Boys have positive covariate values and girls have negative covariate values). The estimated standard deviations of the random signals are denoted as $\widehat{\sigma_F}$. The estimated standard deviations of the random noise in different data sets are denoted as $\widehat{\sigma_E}$.}
\label{tab:SIPCA-A-Growth1}
\begin{center}
\begin{tabular}{cccc}
Component &  $\widehat{\beta}$ & $\widehat{\sigma_F}$ & $\widehat{\sigma_E}$ \\
\hline
    Joint 1    & 0.00  &  7.30  &            \\
    Joint 2    &    10.01  &  5.57  &             \\
    Amplitude 1     &   -4.31  &  3.95  &  0.15     \\
    Amplitude 2     &         0  &  2.51  &  0.15     \\
    Phase 1   &         0  &  5.19  &  0.41     \\
    Phase 2   &         0  &   4.7  & 0.41      \\
    Phase 3   &   -4.37  &  3.04  & 0.41      \\
    Phase 4   &         0  &  2.13  & 0.41      \\
\end{tabular}
\end{center}

\end{table}

In this example, sex is the only covariate. The mean effect of sex is non-trivial for a joint component and one for each individual component, as shown in Table~\ref{tab:SIPCA-A-Growth1}. To highlight the  sex effect on pediatric growth patterns, the components explaining most of the sex effect are shown in Figure \ref{fig:SIPCA}.
In particular, we present the second joint component, the first amplitude component, and the third phase component.
It is well known that pubertal growth spurt appears differently in boys and girls (both the intensity and the timing).
The amplitude component plot and the phase component plot in the upper middle and right panels of Figure \ref{fig:SIPCA} confirm this.
Girls (with negative covariates and negative coefficients for both components) tend to have a lower growth rate peak at a younger puberty age than boys.
The joint component plot in the upper left panel shows an interesting joint pattern of amplitude and phase:  girls (with negative covariates and a positive coefficient) tend to have a smaller and shorter gap between pre-pubertal dip and the pubertal peak compared to boys. The conventional FPCA fails to capture these patterns.

Some scatter plots for the conditional expectations of the SIFA factors are shown in the bottom panels of Figure \ref{fig:SIPCA}.
The scores are roughly elliptically distributed, indicating that there is no obvious association between different components.
Therefore, when used in further statistical analyses such as prediction and inference, the SIFA factors may yield better results than the FPCA scores.

\end{document}